\documentclass[pra,a4paper,aps,twocolumn,superscriptaddress,longbibliography]{revtex4-2}

\usepackage[english]{babel}

\usepackage{graphicx} 
\usepackage{wrapfig}
 
\usepackage{pgfplots}
\pgfplotsset{compat=1.18}
\usepgfplotslibrary{groupplots}

\usepackage[caption=false]{subfig}

\usepackage{amsmath,amssymb}
\usepackage{lipsum}
\usepackage{physics} 

\usepackage[all]{xy}
\usepackage{qcircuit} 

\DeclareFontFamily{U}{wncy}{}
\DeclareFontShape{U}{wncy}{m}{n}{<->wncyr10}{}
\DeclareSymbolFont{mcy}{U}{wncy}{m}{n}
\DeclareMathSymbol{\Sh}{\mathord}{mcy}{"58}

\usepackage[colorlinks=true,linkcolor=blue,citecolor=blue,urlcolor=blue]{hyperref}
\usepackage[capitalise]{cleveref}

\usepackage{comment}

\newcommand{\catgk}[1][\alpha]{g^{(#1)}_k(\boldsymbol{t})}
\newcommand{\fockgk}[1][n]{f^{(#1)}_k(\boldsymbol{t})}

\date{\today}

\begin{document}

\title{Improved GKP magic states from error-corrected non-Gaussian quantum states}
\author{Sharon David}
\author{Jack Davis}
\author{Ulysse Chabaud}
\author{Francesco Arzani}
\email{francesco.arzani@inria.fr}
\affiliation{QAT Team, DIENS, École normale supérieure, PSL University,\\
CNRS, INRIA, 45 rue d'Ulm, Paris 75005, France \\}

\begin{abstract}

Gate teleportation, together with magic state distillation, is a promising route towards fault-tolerant, universal computation. In the context of bosonic quantum computation, Baragiola et al.\ (2019)~\cite{Baragiola_2019} showed that within the framework of Gottesman--Kitaev--Preskill codes, encoded magic states suitable for distillation can be produced by error correcting Gaussian states, such as the vacuum. Here, we show that applying the same framework to simple non-Gaussian input states can significantly improve the quality of the magic states obtained, reducing the overall resources for the complete distillation procedure.  We focus on superpositions of coherent states or Fock states, showing that many can lead to improvements in the generation of high-quality encoded magic states, which in some cases reduces the resources required for magic state distillation by about a factor of $3$. We also investigate the primary source of these improvements and find that, unlike what was previously conjectured, the suitability of input states is not fully explained by symmetry arguments. Instead, the best states seem to avoid projection near stabilizer states as a result of the error correction procedure.
\end{abstract}

\maketitle

\section{Introduction}

Quantum error correction is one of the keys to unlocking the advantages of quantum computation~\cite{shor_scheme_1995}. In the discrete-variable (DV) paradigm, a dichotomy exists between Clifford and non-Clifford operations. The former are one of the most important tools for stabilizer codes, and the backbone of numerous protocols. However, their action on stabilizer states can be simulated efficiently on a classical computer~\cite{gottesman1998heisenberg}, so a non-Clifford resource is needed to achieve quantum advantage, and is sufficient to promote Clifford computation to a universal model. 

When used to protect computation, a quantum error-correction scheme must include instructions to compile logic gates in a fault-tolerant way. One sufficient condition for achieving this is for a gate to admit a transversal implementation, which essentially prevents errors from propagating too rapidly during the computation.
Unfortunately, the Eastin--Knill theorem~\cite{Eastin_2009} prevents any quantum error-correcting code from having a universal set of transversal gates. A solution to this problem is to use gate teleportation~\cite{Gottesman_Chuang_1999}. This protocol consumes a fixed resource auxiliary state, often a non-stabilizer state called a magic state, to realize a non-Clifford gate on an unknown input state. Resource states can be refined offline via magic state distillation, which is a set of protocols built from Clifford operations and Pauli measurements that consumes many noisy copies of a resource state to obtain fewer states of better quality, and can iteratively produce resource states that are arbitrarily close to pure states~\cite{Bravyi_2005}.

These techniques can in principle be applied at the logical level regardless of the underlying physical nature of the information carriers. Continuous-variable (CV) systems in particular, also known as quantum systems with bosonic degrees of freedom, have become especially attractive for carrying quantum computations~\cite{Braunstein2005,Andersen2010,Weedbrook2012,Adesso2014,Serafini2023,Matsuura_Menicucci_Yamasaki_2026}. This is in part due to the multitude of quantum systems naturally described by continuous variables, together with the rapidly developing experimental control of these systems. Such platforms include photonics~\cite{Pfister2020}, trapped ions~\cite{Andersen2015}, and superconducting circuits~\cite{Blais2021,Ma2021control}. Regardless of the physical platform, a bosonic code is used to encode discrete logical information into an infinite-dimensional bosonic Hilbert space. One such code that has attracted significant attention is the Gottesman--Kitaev--Preskill (GKP) code, known for performing well against small displacement errors and pure loss~\cite{Gottesman_2001, Albert_performance_bosonic_codes_2018, Harris_2025}. Experimental preparation and control of code states has been demonstrated in each of the above-mentioned platforms \cite{Konno_optical_GKP_2024, larsen_integrated_2025, Fluhmann_Home_2019, matsos_universal_2025, fontbote_Schmidt_EC_of_beamsplitter_qunaught_2026, campagne-ibarcq_quantum_2020, NordQuantique_autonomous_GKP_2024}, including demonstrations where error-corrected encoded states exhibit longer coherence times than unencoded ones~\cite{Sivak_2023, NordQuantique_autonomous_GKP_2024}.

While a key feature of the GKP code is that encoded Clifford operations can be implemented with the relatively accessible Gaussian operations, there is still the problem of performing logical non-Clifford gates. To this end, Ref.~\cite{Baragiola_2019} introduced a protocol to produce high-quality logical magic states based on the observation that logical distillable magic states can be obtained by error-correcting certain Gaussian states. In particular, among thermal states, the vacuum was found to yield the best distillable magic states on average, while an increase in temperature yields a degradation in quality. Given a source of low-temperature thermal states, the resulting noisy logical magic states would then be fed into a concatenated distillation code to obtain a higher quality logical magic state, i.e.,~one with a purity sufficient to inject a logical non-Clifford gate of desirable quality. Altogether this protocol demonstrates that encoded Clifford operations can be made universal while remaining fault-tolerant, provided a supply of logical stabilizer states (needed for the error correction) and low-temperature thermal states~\cite{Baragiola_2019}. 

While the above scheme leads to fault-tolerant universality, the distillation step incurs a significant overhead, which motivates alternative strategies. One such strategy, in the context of photonic platforms, is to forgo both the distillation procedure and the generation of stabilizer states and instead focus only on preparing high-quality encoded magic states directly \cite{Yamasaki_cost_reduced_2020}.
While a supply of high-quality approximate logical magic states is sufficient for encoded universality, current experimental realizations are still far from the required quality \cite{Konno_optical_GKP_2024,larsen_integrated_2025}. Moreover, in practice the GKP code will be concatenated with a qubit-level code and it is unclear how the resources required for logical magic state preparation will scale in this scenario. This motivates investigating the use of non-Gaussian states that are easier to prepare from an experimental standpoint.
In this work, we determine how the introduction of simple non-Gaussian states at the beginning of the protocol from Ref.~\cite{Baragiola_2019} can reduce the total amount of resources (input states) required to approximately prepare a target GKP magic state.

In addition to the operational motivation of reducing the resources required for GKP quantum computation, our approach is also of theoretical interest from the perspective of general resource conversion.  In particular, the protocol from \cite{Baragiola_2019} was initially surprising since GKP-encoded universality was unlocked using Gaussian states, which are conventionally thought of as classical or ``quasi-classical'' resources in CV systems.  This can be seen as an instance of a free object within one resource theory (i.e.\ non-Gaussianity) becoming a resourceful object in a different resource theory (i.e.\ encoded GKP magic); see \cite{Yamasaki_cost_reduced_2020, Pantaleoni_SSD_PRL_2020, GKP_magic_PRL_2022, Calcluth_vacuum_provides_2023, Shaw_2024, Calcluth_Sufficient_Condition_PRXQ_2024, Bridging_magic_2025, Davis2026, Hosseinynejad_Realistic_GKP_are_universal_PRL_2026} for more in the context of the GKP code and \cite{Descamps_superselection_and_bosonic_resources_PRL_2024, descamps_heisenberg_weyl_bosonic_phasespaces_arxiv_2025, Descamps_optica_2026, Deneris_free_vs_resource_2026} for more general discussions.  Our work contributes to this growing research program by investigating the subtle relationship between non-Gaussianity in the input continuous-variable state and the discrete-variable non-stabilizerness of the output logical state.

Concretely, we show that applying GKP error correction to cat states and (superpositions of) Fock states can significantly improve the average fidelity with distillable magic states with respect to error-corrected Gaussian states. We focus on these two families of states for two main reasons:
first, they can be easily prepared with the same platforms used to engineer
GKP states~\cite{Sivak_2023,CampagneIbarcq2020,Fluhmann2019,deNeeve2022,Vlastakis2013,Hofheinz2009}, sometimes as intermediate byproducts~\cite{Tzitrin2020,Eaton2024}; second, their structure considerably simplifies the analytical calculations. The improvement is quantified by the probability of obtaining distillable logical states. We study both fixed non-Gaussian states, as well as optimized superpositions of these non-Gaussian states, and analyze the properties of the states that offer good performances, in particular their probability of being projected on different regions of the logical Bloch sphere after GKP error correction. We identify one potential explanation for good performance to be that ``good'' states have low probabilities of being projected into the vicinity of stabilizer states. This contrasts with the natural intuition that good states should primarily have high probabilities of landing close to the magic states themselves, for example because they share the same symmetries~\cite{Conrad_thesis_2024}.

We then turn to the practical implications: we study whether the observed improvements in success probability lead to noticeable advantages at the distillation stage. We evaluate the number of CV resources required to obtain magic states exceeding a given target fidelity via distillation. We show that starting from superpositions of coherent or Fock states can significantly reduce the number of error correction circuits that need to be executed, hereafter referred to as \textit{distillation cost}. In fact, we find specific optimized superpositions where the distillation cost is about a third of that of the vacuum. 

The rest of this paper is organized as follows. In~\cref{sec: prelims} we present the preliminaries required to understand the results and explain the concepts of magic states, distillation and the GKP code. We also review the error correction gadget by which one obtains GKP logical states from arbitrary CV states. In~\cref{sec: primary-results} we derive the analytical expressions used to compute the description of the states obtained when error-correcting cat states and Fock states. We also include the probability of obtaining a given minimum fidelity with magic states. \cref{sec: practical_advntgs} shows the practical distillation improvements when working with cat and Fock states, and~\cref{sec: output_distribs} shows the output distributions of these input states on the Bloch sphere; here, we identify the features that make a state more resourceful after error correction. \cref{sec: conc} concludes by summarizing the results and highlights some open questions.

\section{Preliminaries} \label{sec: prelims}
\subsection{Magic states and their distillation} \label{subsec: msd}
\begin{figure}
    \centering
    \includegraphics[width=0.6\linewidth]{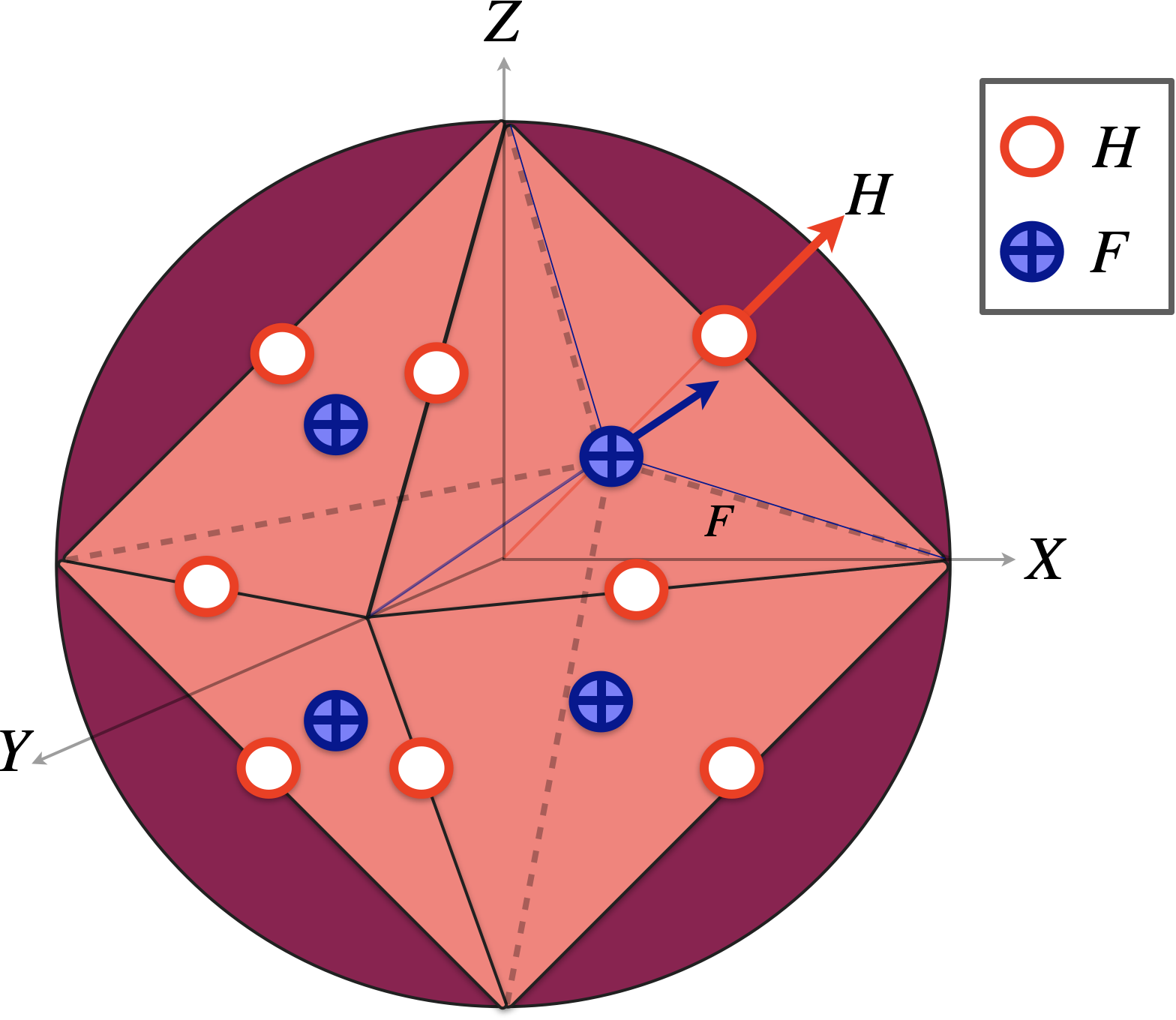}
    \caption{The Bloch sphere is a geometric representation of a qubit. Its antipodal points correspond to the 6 Pauli eigenstates, and their convex hull forms the stabilizer octahedron. There exist twelve $H$-type and eight $F$-type pure non-stabilizer ``magic'' states that are projections onto the surface of the sphere from the centers of the edges and faces of the octahedron, respectively.}
    \label{fig:bloch_sphere_magic_states}
\end{figure}

The pure stabilizer states for a single qubit are the six eigenstates of the Pauli operators $\hat X$, $\hat Y$, and $\hat Z$.  Their convex hull forms an octahedron 
known as the stabilizer polytope, and any mixed state within this polytope is called a stabilizer state; see~\cref{fig:bloch_sphere_magic_states} for the Bloch sphere representation. The single-qubit \textit{Clifford group}, i.e., the group of
unitaries $\hat{C}$ (modulo global phases) such that
$\hat{C}\hat{\sigma}^\mu\hat{C}^\dagger \in \mathcal{P}_1$ for
all Pauli operators
$\hat{\sigma}^\mu \in \mathcal{P}_1 = \langle i\hat{I}, \hat{X},
\hat{Y}, \hat{Z}\rangle$, is isomorphic to the group of symmetries
of the stabilizer polytope, mapping stabilizer states to stabilizer
states. Non-stabilizer, or \textit{magic} states, are those not contained within this stabilizer polytope.  There are two species of pure magic states that deserve special attention: $H$-type and $F$-type~\cite{Bravyi_2005, Kubischta_Ftype}~\footnote{Note that $F$-type magic states are called $T$-type in~\cite{Bravyi_2005}. We use the notation $F$ here to avoid confusion, since $H$-type states can be used to implement what has come to be known as $\hat{T}$ gate, or $\pi/8$-gate in the literature.}. These are eigenstates of Clifford operators, and can thus be seen as the set of pure states that are normal to the centres of the edges and faces of the stabilizer octahedron, respectively. As such, there are twelve $H$-type states and eight $F$-type states. States within one class can be reached from any representative of the class under Clifford operations. In other words, each class is the Clifford orbit of any of its representatives. The canonical representative of each orbit is
\begin{equation}
\begin{aligned}\label{eqn:H_plus_state}
    \ket{+H} &= \cos\left(\frac{\pi}{8}\right)\ket{0} + \sin\left(\frac{\pi}{8}\right)\ket{1} \quad \text{and} \quad \\
    \ket{+F} &= \cos(\beta)\ket{0} + \sin(\beta)\ket{1}, \hspace{0.5em} \cos(2\beta) = \frac{1}{\sqrt{3}},
\end{aligned}
\end{equation}
which corresponds to the $+1$ eigenstates of the Clifford unitaries
\begin{equation}
\begin{aligned}
    H &= \frac{1}{\sqrt{2}} \begin{pmatrix}
        1 & 1 \\ 1 & -1
    \end{pmatrix} \quad \text{and} \quad
    F = \frac{e^{i \pi/4}}{\sqrt{2}} \begin{pmatrix} 1 & 1 \\ i & -i\end{pmatrix}.
\end{aligned}
\end{equation}
These states are of key importance because they can be consumed via gate-teleportation to produce non-Clifford gates up to Clifford corrections, which then promotes stabilizer circuits to quantum universality~\cite{Gottesman_1999}.
The main result from~\cite{Bravyi_2005} is that \textit{magic state distillation} can be used to probabilistically convert many copies of magic states to a single copy with arbitrarily high fidelity with an $H$- or $F$-type pure state, provided the initial states are close enough to any state of the respective class. Crucially, magic state distillation schemes can be constructed using only Clifford operations and Pauli measurements~\cite{Bravyi_2005, Reichardt_2005, Reichardt_MSD_2006, campbell2009structureprotocolsmagicstate, Litinski_2019, xu2026distillingmagicstatesbicycle}. A general $n$-to-$1$ procedure to distill a specific magic state $\ket{m}$ can always be thought of as follows: take as input $n$ noisy copies of $\ket{m}$, $\rho_{\tilde{m}}^{\otimes n}$, project them onto the codespace of some stabilizer quantum error-correcting code (by extracting the syndrome and post-selecting on the trivial syndrome), un-encode the result to the first register, then trace out the remaining $n-1$ qubits~\cite{campbell2009structureprotocolsmagicstate}. If the fidelity  $F_{\text{in}} =\langle m | \rho_{\tilde{m}}|m\rangle$  is above a certain value $F_\mathrm{th}$, called the \textit{distillation threshold}, which depends on the chosen error-correcting code~\cite{Bravyi_2005}, then the result is the probabilistic generation of a single qubit state $\bar{\rho}_{\tilde{m}}$ that will have a higher fidelity to the target state than $\rho_{\tilde{m}}$. This process can then be iterated with several copies of the output state, $\bar{\rho}_{\tilde{m}}^{\otimes n}$, now as input, until a desired fidelity with $\ket{m}$ is achieved. Finally, the state thus obtained is consumed to implement a non-Clifford gate on the qubits. The quality of such a non-Clifford gate is directly related to the quality of the final magic state. As a result, for a given desired accuracy, many input states can be required to realize the needed distillation iterations, potentially creating large overheads. This highlights the importance of producing states $\rho_{\tilde{m}}$ with sufficiently high initial fidelity to the target magic state. 

In this work we benchmark against the 15-to-1 qubit Reed--Muller distillation code~\cite{Bravyi_2005}. As the name suggests, this procedure consumes $15$ noisy magic states to distill $1$ magic state closer to the target class. While more resource-intensive than other codes, this code was chosen over smaller codes because it offers cubic error suppression per round of distillation~\cite{howard_small_2016}.

\subsection{Continuous-variable quantum mechanics }
We work with physical systems with bosonic degrees of freedom, modelled as quantum harmonic oscillators, each with bosonic operators $\hat a$ and $\hat a^\dagger $ such that $\left[\hat a , \hat a ^\dagger \right]=1$~\cite{ulf_leonhardt_text}. Position and momentum operators are defined as, respectively,
\begin{equation}
    \hat q = \frac{\hat a + \hat a^\dagger}{\sqrt{2}} , \quad \hat p = \frac{\hat a - \hat a^\dagger}{i\sqrt{2}} .
\end{equation}
With this convention, $[\hat q, \hat p] = i$ with $\hbar = 1$. The (non-normalizable) eigenstates of the position and momentum operators are denoted $\ket{s}_q$ and $\ket{t}_p$, respectively. 
Shifts in position and momentum can be represented with displacement operators
\begin{equation} \label{eqn:displacement_op}
    \hat D(\alpha) = e^{\alpha \hat{a}^\dagger -\alpha^* \hat{a}},
\end{equation}
where $\alpha \in \mathbb{C}$, and is defined as $\alpha = \frac{1}{\sqrt{2}}(q_0 + i p_0)$.
Coherent states are the eigenstates of the annihilation operator $\hat{a}\ket{\alpha} = \alpha\ket{\alpha}$, with complex eigenvalue $\alpha$. Coherent states can also be defined operationally as displaced (\cref{eqn:displacement_op}) vacuum states 
\begin{equation}
    \hat{D}(\alpha) \ket{0} = \ket{\alpha}.
\end{equation} We can give a similar operational definition for Fock states, which are eigenstates of the number operator $\hat{n}\ket{n} =\hat{a}^\dagger\hat{a}\ket{n} = n\ket{n} $. These states are also obtained as
\begin{equation}
    \ket{n} = \frac{1}{\sqrt{n!}}\left(\hat a^\dagger \right)^n \ket{0},
\end{equation}
where $\ket{0}$ is the vacuum or ground state of the number operator. Finally, in order to define Gaussian states, it is convenient to introduce Wigner functions, which represent states and operators in phase space. For an operator $\hat{A}$ satisfying appropriate regularity conditions, the Wigner function can be defined as~\cite{ulf_leonhardt_text}
\begin{equation}
    W_{\hat A}(q,p) = \frac{1}{2 \pi} \int_{-\infty}^\infty  e^{i p x} \bra{q - \frac{x}{2}} \hat A \ket{q + \frac{x}{2}} dx.
\end{equation}
Gaussian states are those whose Wigner function is Gaussian
\begin{equation}
    W_G(\xi) = \frac{1}{\sqrt{(2 \pi)^n \det \Gamma}} \exp \left ( -\frac{1}{2} (\xi - \xi_0)^T \Gamma^{-1} (\xi - \xi_0)\right),
\end{equation}
where $\xi^T = (\boldsymbol{q}^T, \boldsymbol{p}^T)$ represents the $2n$ canonical variables. The first moments $\xi_0$ and the covariance matrix $\Gamma$ uniquely specify any Gaussian state $G$.

\subsection{The GKP code}
GKP codes are a class of codes based on translational invariance, encoding a $d$-dimensional qudit into $n$ bosonic modes~\cite{Gottesman_2001}. Ideal code states are represented as infinite sums of Dirac delta distributions in either the position or momentum basis. Here we focus on the original case of encoding a qubit into a single mode (i.e.\ $d=2$ and $n=1$), where the encoded computational basis is given by 
\begin{equation} \label{eqn:gkp_logicals} 
\begin{aligned}
    \ket{0_{\text{L}}} &=\frac{1}{2\sqrt[4]{\pi}}\sum_l \ket{2l\sqrt{\pi}}_{q} \\
    \ket{1_{\text{L}}} &= \frac{1}{2\sqrt[4]{\pi}}\sum_l \ket{(2l + 1)\sqrt{\pi}}_{q},
\end{aligned}
\end{equation}
and $1/2\sqrt[4]{\pi} $ is a normalization constant chosen so that the Wigner function of code states is normalized on the unit cell of the logical (square) lattice $\left[0, 2\pi \right)^2$~\cite{Baragiola_2019}. Indeed, ideal code states are  unphysical, and not normalizable in the usual sense. However, approximate code states can be defined~\cite{Matsuura_equivalent_approximate_2020} and prepared experimentally, and are sufficient for quantum error correction and encoded computation~\cite{Hosseinynejad_Realistic_GKP_are_universal_PRL_2026, Matsuura_Menicucci_Yamasaki_2026}. Here we focus on ideal code words \eqref{eqn:gkp_logicals}, which are equivalently defined as the simultaneous +1 eigenstates of the GKP stabilizers. Since this code was built to correct small shifts in phase space, the stabilizers of the GKP code are naturally the displacement operators
\begin{equation}
    \hat S_q = e^{-i 2 \sqrt{\pi} \, \hat{p}}, \quad \hat S_p = e^{i  2 \sqrt{\pi} \,\hat{q}}.
\end{equation}
These operators generate the GKP stabilizer group: a set of commuting displacement operators ($[\hat S_q, \hat S_p] = 0$) parametrized by the lattice $2\sqrt{\pi}\, \mathbb Z \times 2\sqrt{\pi}\, \mathbb Z \subset \mathbb R^2$ within phase space~\cite{Conrad_2022}.  
The logical Pauli operators are instead associated with $\sqrt{\pi}$-shifts in either position or momentum space,
\begin{equation} \label{eqn:gkp_logical_X_Z}
    \hat{X}_{\text{L}} = e^{- i \sqrt{\pi} \, \hat{p}}, \qquad \hat{Z}_{\text{L}} = e^{i \sqrt{\pi} \,\hat{q}}.
\end{equation}
These satisfy the required Pauli commutation relation $\hat X_{\text{L}} \hat Z_{\text{L}} = - \hat Z_{\text{L}} \hat X_{\text{L}}$ and commute with the stabilizer displacements. These are physical unitary operators that act on the Hilbert space of the oscillator, but they are not Hermitian. Instead, Hermitian bosonic operators that act as Pauli operators on the code space can be defined using the encoded computational basis \eqref{eqn:gkp_logicals} as
\begin{equation}\label{eqn:gkp_logical_paulis_sigma_mu}
    \hat{\sigma}_{\text{L}}^\mu = \sum_{j,k=0}^1 \sigma_{jk}^\mu \ketbra{j_{\text{L}}}{k_{\text{L}}},
\end{equation}
where $\sigma^\mu_{jk}$ for $\mu = \{0,1,2,3\}$ is the $(j,k)$-th element of the $\sigma^\mu$ Pauli matrix~\cite{Baragiola_2019}. The encoded identity operator, $\hat \sigma_{\text{L}}^0$, is also the projector onto the GKP code space within the bosonic Hilbert space:
\begin{equation} \label{eqn:gkp_projector}
    \hat{\Pi}_{\text{GKP}} = \hat{\sigma}_{\text{L}}^0 = \ketbra{0_{\text{L}}}+\ketbra{1_{\text{L}}}.
\end{equation}

\subsection{Distillable magic states through GKP error correction}

\begin{figure}
    \centering
   \[
\Qcircuit @C=0.9em @R=1.5em {
\lstick{\rho_\mathrm{in}} & \ctrl{1} & \qw & \gate{\hat X(-t_q )}
  & \targ & \qw & \gate{\hat Z(-t_p )}
  & \rstick{\rho_\text{out}} \qw \\
\lstick{\ket{0_L}} & \control \qw & \measureD{\hat{p}}
  & \dstick{t_q} \cw \cwx[-1] & & & & \\
\lstick{\ket{0_L}} & \qw & \qw & \qw & \ctrl{-2}
  & \measureD{\hat{p}} & \dstick{t_p} \cw \cwx[-2]  &
}
\]

    \caption{The GKP error-correction gadget corrects any arbitrary input state and projects it into the logical GKP code space.}
    \label{fig:gkp_error_correction_gadget}
\end{figure}
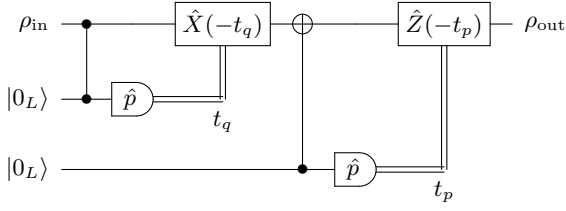

Displacement noise applied to code states can drive them outside the logical space. Such noise can be determined up to logical corrections by measuring the stabilizers, i.e.,\ extracting the error syndrome. A subsequent correction can then undo the error by negatively displacing the qubit back into the GKP code space~\cite{Gottesman_2001}; see~\cref{fig:gkp_error_correction_gadget} for the circuit depiction of this process.  Here the error correction for position and momentum are performed separately.  First, for the $q$ quadrature, the input state is entangled with an ideal GKP ancilla in the logical zero state via a $\hat{C}_Z = e^{i \hat{q}\otimes \hat{q}}$ gate.  Then the momentum of the ancilla is measured, e.g.,\ via homodyne detection.  The  outcome is denoted $t_q$ and a corresponding displacement $\hat X(-t_q ) = e^{- i (-t_q) \hat{p}}$ is applied to the data mode. 
Similarly, the input is then entangled with the second ancilla and measured. The syndrome value is denoted $t_p$ for the displacement in the $p$  quadrature, and it is corrected with a $\hat Z(-t_p ) = e^{i (-t_p)\hat{q}}$ operation. 

The overall operation can be summarized as a Kraus operator parametrized by $\boldsymbol{t} = (t_q, t_p) \in \mathbb{R}^2$
\begin{equation} \label{eqn:gkp_error_corr_kraus_op}
    \hat{K}_{\text{EC}}(\boldsymbol{t}) = \hat{K}^p_{\text{EC}}(t_p) \hat{K}^q_{\text{EC}}(t_q) = \hat{\Pi}_\mathrm{GKP} \hat{V}(\boldsymbol{-t}),
\end{equation}
where
\begin{equation}
\begin{aligned}
\hat{K}^q_{\text{EC}}(t_q) &= \Sh_{\sqrt{\pi}}(\hat{q})\hat{X}(-t_q) \\
\hat{K}^p_{\text{EC}}(t_p) &= \Sh_{\sqrt{\pi}}(\hat{p})\hat{Z}(-t_p),
\end{aligned}
\end{equation}
and
\begin{equation} \label{eqn:gkp_displacement_operator}
     \hat{V}(\boldsymbol{-t}) = \hat{Z}(-t_p)\hat{X}(-t_q).
\end{equation}
Here $\Sh_{\sqrt{\pi}}(s) = \sum_{n \in \mathbb{Z}} \delta(s - \sqrt{\pi}n)$ denotes a Dirac comb with spacing $\sqrt{\pi}$ and overall shift $s$, so that $\Sh_{\sqrt{\pi}}(\hat q) = \sum_{n \in \mathbb{Z}} \ket{\sqrt{\pi}n}\!\bra{\sqrt{\pi}n}_{q}$.  

The main observation of Ref.~\cite{Baragiola_2019} is that this procedure is relevant for states beyond displaced code states: \textit{any} bosonic state will result in a logical state.  When~\cref{eqn:gkp_error_corr_kraus_op} acts on an input state $\rho_\mathrm{in}$, the state is first displaced by a random amount depending on the measurement outcome via~\cref{eqn:gkp_displacement_operator}, with the probability of each outcome depending on the state, and is then projected onto the GKP code space with~\cref{eqn:gkp_projector}.

\subsubsection{Computing Bloch vector coefficients}\label{subsec: error_correction_method}
The Kraus operator acting on an input state produces a syndrome-dependent, unnormalized output state \begin{equation}\label{eqn:rho_out}
    \hat{\bar \rho}_\mathrm{out}(\boldsymbol{t}) = \hat{K}_{\text{EC}}(\boldsymbol{t}) \hat\rho_\mathrm{in} \hat{K}_{\text{EC}}^\dagger(\boldsymbol{t}),
\end{equation}
where the bar reminds us that it is some logical (mixed) state. Being such an ideal code state, it can be normalized by dividing by the probability density function \begin{equation}
\bar{r}_0\left(	\boldsymbol{t}\right) = \text{Tr}[\hat{V}(\boldsymbol{-t})\hat{\rho}_\mathrm{in}\hat{V}^\dagger(\boldsymbol{-t})\hat{\Pi}_\mathrm{GKP}].
\end{equation}
The full syndrome-dependent logical qubit state can be written in terms of the logical Pauli operators
\begin{equation}
    \hat \rho_\mathrm{out}(\boldsymbol{t}) = \frac{1}{2}[r_0(\boldsymbol{t}) \hat \sigma_L^0 + r_1(\boldsymbol{t}) \hat \sigma_L^1 + r_2(\boldsymbol{t}) \hat \sigma_L^2 + r_3(\boldsymbol{t}) \hat \sigma_L^3],
\end{equation}
where the logical Bloch vector coefficients
\begin{equation}\label{eq:bloch_4_vector}
    \vec{r}(\boldsymbol{t}) = (r_0(\boldsymbol{t}), \underbrace{r_1(\boldsymbol{t}), r_2(\boldsymbol{t}), r_3(\boldsymbol{t})}_{\vec r_B(\boldsymbol t)=\text{Bloch 3-vector}})
\end{equation}
can be found by calculating the overlap of the error-corrected output state $\hat{\bar \rho}_\mathrm{out}$ with the GKP logical Pauli operators~\cite{Baragiola_2019}
\begin{align}
    \bar r_\mu(\boldsymbol{t}) = \text{Tr}[\hat {\bar\rho}_\mathrm{out}(\boldsymbol{t}) \hat \sigma_L^\mu] & = \text{Tr}[\hat{\Pi}\hat{V}(\boldsymbol{-t})\hat{\rho}_\mathrm{in}\hat{V}^\dagger(\boldsymbol{-t})\hat{\Pi}\hat{\sigma}^\mu_L]\\
    & = \text{Tr}[\hat{V}(\boldsymbol{-t})\hat{\rho}_\mathrm{in}\hat{V}^\dagger(\boldsymbol{-t})\hat{\sigma}_L^\mu] \label{eqn:overlap}.
\end{align} The normalized Bloch vector coefficients of~\cref{eq:bloch_4_vector} are obtained as \begin{equation}\label{eqn:r_overlap_normalization}
 r_\mu(\boldsymbol{t}) = \frac{\bar r_\mu(\boldsymbol{t})}{\bar r_0(\boldsymbol{t})}.
\end{equation} 
This procedure can also be carried out with the state-vector formalism when working with pure states:
\begin{equation} \label{eqn:r_overlap_sv_method}
    \bar r_\mu(\boldsymbol{t}) = \sum_{j k } \sigma_{jk}^\mu \bra{k_L}\hat{V}(-\boldsymbol{t})\ketbra{\psi}\hat{V}^\dagger(-\boldsymbol{t})\ket{j_L}.\\
\end{equation}

\subsubsection{Best fidelity}
Let us define $\ket{+H} $ as the $+1$ eigenstate of the Hadamard operator $\hat H$. The logical fidelity of the output state with respect to the $\ket{+H_L}$ logical magic state is \begin{equation} \label{eqn: best_fidelity_H}
    F  = \bra{+H_L} \hat{\rho}_\mathrm{out}(\boldsymbol{t}) \ket{+H_L} = \frac{1}{2}\left[1 + \vec{r}_H \cdot \vec{r}_B(\boldsymbol{t})\right],
\end{equation}
where $\vec{r}_H \cdot \vec{r}_B(\boldsymbol{t})$ is the ordinary scalar product between the Bloch 3-vectors of the Hadamard eigenstate and the logical state obtained after error correction. 
We are interested in the best fidelity between the output state and \textit{any} pure magic state of either class, therefore it is necessary to maximize over the Clifford orbit
\begin{equation}
\begin{aligned}
    F_{H,\mathrm{best}}(\boldsymbol{t}) &= \max_{ \hat{C} \in \mathrm{Clifford}}\; F \left(\hat{\rho}_\mathrm{out}(\boldsymbol{t}), \hat{C}  \ket{H} \right), \\
 	F_{F,\mathrm{best}}(\boldsymbol{t}) &= \max_{ \hat{C} \in \mathrm{Clifford}} \; F \left(\hat{\rho}_\mathrm{out}(\boldsymbol{t}), \hat{C}  \ket{F}  \right).
\end{aligned}
\end{equation}
Here the optimization over the Clifford group is done via brute force. That is, the dot product in~\cref{eqn: best_fidelity_H} is computed between the relevant state's Bloch vector and each of the 12 $H$-states' Bloch vectors; the maximal value corresponds to the state with the best fidelity.

\subsubsection{Success probability}
The above details how to compute the \textit{best fidelity} for each syndrome outcome $\boldsymbol{t}$. Individual outcomes have zero probability, so the expected performance is instead measured via the \textit{success probability} of obtaining \textit{at least} a given value for the fidelity $F^*$, integrated over all possible syndrome outcomes~\cite{Baragiola_2019}
\begin{equation}
\begin{split}
    P_\mathrm{success} = \int_{\boldsymbol{t}:F_{H,\mathrm{ best}}(\boldsymbol{t})\geq F^*} \bar{r}_0\left(\boldsymbol{t}\right) d^2 t.
\end{split}
\end{equation}
To facilitate numerical integration, this equation can also be written using an indicator function $\chi(\boldsymbol{t})$ as
\begin{equation}\label{eqn:success_probability}
    P_\mathrm{success} = \int_{\boldsymbol{t} \in [-\sqrt{\pi}, \sqrt{\pi}]^2} d\boldsymbol{t} \bar{r}_0(\boldsymbol{t}) \chi(\boldsymbol{t}),
\end{equation}
where 
\begin{equation}
\chi(\boldsymbol{t}) = 
\begin{cases}
    1 & F_{H,\mathrm{best}}(\boldsymbol{t}) \geq F^*\\
    0 & F_{H,\mathrm{best}}(\boldsymbol{t}) < F^*.
\end{cases}
\end{equation}
Once more, this measure can be adapted for the $F$-type states, simply by replacing $F_{H,\mathrm{best}}(\boldsymbol{t})$ with $F_{F,\mathrm{best}}(\boldsymbol{t})$. 

This success probability measure is similar to the quantity mentioned in \cite{Calcluth_Sufficient_Condition_PRXQ_2024} but the crucial difference is that there the fidelity with the orbit of magic states is taken \textit{after} averaging the output state over the syndrome probabilities. Instead, here we assume that a Clifford correction can be applied depending on the syndrome, then we compute the fidelity to the closest magic state in the orbit, and finally we average this quantity over all syndromes. This can be achieved with active feedback or by adapting the Pauli frame. 
In essence, we are performing an average over the collection of normalized post-selected states, similar to~\cite{Baragiola_2019} (which is restricted to input thermal states). As a consequence, the values we find for the success probability are always greater than or equal to those in~\cite{Calcluth_Sufficient_Condition_PRXQ_2024}.

\section{GKP Error-Correction with Non-Gaussian Input States} \label{sec: primary-results}
In this section we compute the output of the circuit in Fig.~\ref{fig:gkp_error_correction_gadget} when the input states are
cat states, Fock states, and superpositions of Fock states. In doing so, we generalize the analytical techniques from~\cite{Baragiola_2019} to accommodate these novel classes of non-Gaussian inputs. We then show that such states can yield relatively higher success probabilities as compared to the vacuum, which was the best input state among those considered in~\cite{Baragiola_2019}.

\subsection{Cat states}

Following common usage in quantum optics,  we call \textit{cat state} \cite{Cochrane_1999} an arbitrary superposition of finitely many coherent states:
 \begin{equation} \label{eqn:gen_cat_state}
    \ket{\Gamma} = \frac{1}{\sqrt{N_\mathrm{cat}}} \sum_{c=1}^\Lambda \bar{\gamma}_c \ket{\alpha_c} \text{ where } \bar{\gamma}_c, \alpha_c \in \mathbb{C},
\end{equation}
where $N_\mathrm{cat}$ is the normalization constant:
\begin{equation}
\begin{split}
    N_\mathrm{cat} & =  \sum_{c, d=1}^\Lambda \bar{\gamma}_c^* \bar{\gamma}_d\bra{\alpha_c }\ket{\alpha_d}\\
    & = \sum_{c, d=1}^\Lambda \bar{\gamma}_c^* \bar{\gamma}_d \exp\!\left[-\frac{1}{2}|\alpha_c |^2 - \frac{1}{2} |\alpha_d|^2 + \alpha_c ^*\alpha_d\right].
\end{split}
\end{equation}
Note that this expression depends on the coefficients $\bar{\gamma}_c$ as well as $\alpha_c$, since coherent states are not orthogonal to each other. $\Lambda$ denotes the number of coherent states in the superposition and is commonly referred to as the number of \textit{legs} of the cat. The simplest example is a superposition of two coherent states, often called a two-legged cat state \cite{PhysRevLett.57.13}. More exotic cat states like the four-legged cat state (or \textit{compass} state), a superposition of four coherent states, have also been prepared in experiments~\cite{Hastrup_2020}.

\subsubsection{Computing Bloch vector coefficients \label{subsubsec:cat_states_bloch_comps}}

By writing the density matrix of a general cat state as
\begin{align}
    \hat{\rho}_{\text{cat}} & = \ketbra{\Gamma}  = \frac{1}{N_\mathrm{cat}} \sum_{c=1}^\Lambda \sum_{d=1}^\Lambda \bar{\gamma}_c \bar{\gamma}_d^* \ketbra{\alpha_c}{\alpha_d} \\
     & = \frac{1}{N_\mathrm{cat}} \left[ \sum_{c=1}^\Lambda |\bar{\gamma}_c|^2 \ketbra{\alpha_c}{\alpha_c} + \sum_{c \neq d}^\Lambda \left( \bar{\gamma}_c\bar{\gamma}_d^* \ketbra{\alpha_c}{\alpha_d} \right) \right], \label{eqn:dm_cat_state_expanded}
\end{align}
we can expand the  Bloch coefficients~\eqref{eqn:r_overlap_sv_method} as
\begin{equation}\label{eqn:cat_state_r_overlap_method} 
\begin{aligned}
    \bar r_{\mu}^\mathrm{(cat)}(\boldsymbol{t}) & =  \mathrm{Tr}[\hat{V}(-\boldsymbol{t}) \hat{\rho}_{\text{cat}}\hat{V}(\boldsymbol{t})\hat{\sigma}^\mu_L]\\
    & = \frac{1}{N_\mathrm{cat}} \sum_{c,d}^\Lambda \bar{\gamma}_c\bar{\gamma}_d^* \mathrm{Tr}[\hat{V}(-\boldsymbol{t}) \ketbra{\alpha_c}{\alpha_d}\hat{V}(\boldsymbol{t})\hat{\sigma}_L^\mu].\\
\end{aligned}
\end{equation}
By using the cyclic invariance of the trace and expanding the logical Pauli operator as in~\cref{eqn:gkp_logical_paulis_sigma_mu} we obtain
\begin{align}
     & \bar r_{\mu}^\mathrm{(cat)}(\boldsymbol{t}) = \frac{1}{N_\mathrm{cat}} \sum_{c, d}^\Lambda \bar{\gamma}_c\bar{\gamma}_d^* \bra{\alpha_d} \hat{V}(\boldsymbol{t}) \hat{\sigma}_{\text{L}}^\mu \hat{V}(-\boldsymbol{t}) \ket{\alpha_c} \nonumber \\
    &= \frac{1}{N_\mathrm{cat}} \sum_{c,d}^\Lambda \bar{\gamma}_c\bar{\gamma}_d^* \Bigg(\sum_{j,k} \sigma_{jk}^\mu \bra{\alpha_d} \hat{V}(\boldsymbol{t})\ket{j_{\text{L}}}\underbrace{\bra{k_{\text{L}}}\hat{V}(-\boldsymbol{t}) \ket{\alpha_c}}_{\catgk} \Bigg), \label{eqn:cat_state_overlap_gk}
\end{align}
We utilize ~\cref{eqn:gkp_logicals}, ~\cref{eqn:gkp_logical_X_Z} and ~\cref{eqn:gkp_displacement_operator}, and apply two operations $_q\!\bra{s}e^{i p_0 \hat{q}} = e^{i p_0 s}{}_q\!\bra{s} $, and $_q\!\bra{q} e^{- i q_0 \hat{p}} = {}_q\!\bra{q - q_0}$ to yield
\begin{align}
    \catgk & = \bra{k_\text{L}}\hat{V}(-\boldsymbol{t})\ket{\alpha_c} = \bra{k_L}e^{i (-t_p) \hat{q}} e^{-i (-t_q) \hat{p}}\ket{\alpha_c} \\
    & =\frac{1}{2\sqrt[4]{\pi}} \sum_l {}_q\!\bra{(2l + k)\sqrt{\pi}}e^{i (-t_p) \hat{q}} e^{-i (-t_q) \hat{p}}\ket{\alpha_c} \\
    & =\frac{1}{2\sqrt[4]{\pi}} \sum_l e^{-i t_p (2l + k)\sqrt{\pi}}\underbrace{{}_q\!\bra{(2l + k)\sqrt{\pi} +t_q} \ket{\alpha_c}}_{\text{coherent state wavefunction}}.\label{eqn:coh_state_overlap}
\end{align}
In the last line, we recognize the wavefunction of the coherent state in the position basis, which is written in general as~\cite{ulf_leonhardt_text}
\begin{equation}
    \psi_\alpha(q) = \pi^{-1/4} \exp[- \frac{(q - q_0)^2}{2} + i p_0 q - \frac{i p_0 q_0}{2}],
\end{equation} with $\alpha = \frac{q_0+ip_0}{\sqrt{2}}$.
We substitute for $q =((2l + k)\sqrt{\pi} +t_q) $ and insert the new wavefunction into ~\cref{eqn:coh_state_overlap} to obtain 
\begin{widetext}
\begin{equation}\label{eqn:cat_gk_expanded}
\begin{split}
    \catgk & = \frac{1}{2\sqrt[4]{\pi}} \underbrace{\frac{\exp{{\left [ k\sqrt{\pi}\left(\frac{-k\sqrt{\pi}}{2} + \sqrt{2}\alpha - T\right) + \frac{1}{2} \left(-i \Im{\sqrt{2}\alpha}(\Re{\sqrt{2}\alpha} - 2\Re{T}) - (\Re{\sqrt{2}\alpha} - \Re{T})^2\right)\right ]}}}{\pi^{1/4}}}_{\text{Gaussian term } G_\alpha(\boldsymbol{t})}\\
    & \times \underbrace{\sum_l \exp \left [ 2 \pi i  \left ( \frac{1}{2}\underbrace{(2i)}_{\Omega_\alpha}l^2  + \left ( \underbrace{ik - \frac{i\sqrt{2}\alpha}{\sqrt{\pi}} + \frac{iT}{\sqrt{\pi}}}_{z_\alpha(\boldsymbol{t})}\right)l \right) \right]}_{\text{Riemann theta function }\Theta(z_{\alpha} (\boldsymbol{t}), \Omega_{\alpha})},\\
\end{split}
\end{equation}
\end{widetext}
where $T = t_q + i t_p$.
This can be notationally condensed to
\begin{equation} \label{eqn:cat_gk}
    \catgk = \frac{1}{2\sqrt[4]{\pi}} G_\alpha(\boldsymbol{t}) \Theta(z_\alpha(\boldsymbol{t}), \Omega_\alpha),
\end{equation}
where $\frac{1}{2\sqrt[4]{\pi}}$ is the normalization constant for the GKP logical state, 
$G_\alpha(\boldsymbol{t})$ is a Gaussian term and $\Theta(z_\alpha(\boldsymbol{t}), \Omega_\alpha)$ is a Riemann theta function \cite{deconinck2004computingriemannthetafunctions} with arguments
\begin{equation}
    z_\alpha(\boldsymbol{t}) =  ik - \frac{i\sqrt{2}\alpha}{\sqrt{\pi}} + \frac{iT}{\sqrt{\pi}} \quad \text{and} \quad \Omega_\alpha = 2i.
\end{equation}
Here, Riemann theta functions are defined as
\begin{equation}\label{eqn:riemann-theta}
    \Theta(z, \Omega) = \sum_{\boldsymbol{n} \in \mathbb{Z}^g} \exp\!\left [ 2 \pi i \left ( \frac{1}{2} \boldsymbol{n}^T\cdot\Omega\cdot \boldsymbol{n} + \boldsymbol{z}^T \boldsymbol{n}\right )\right],
\end{equation}
where $z \in \mathbb{C}^g$ and $\Omega \in \mathbb{H}_g$. $\mathbb{H}_g$ is the Siegel upper-half space of all complex and symmetric ($\Omega^T = \Omega$), $g \times g$ matrices with a positive-definite imaginary part. Here $g$ represents the genus, which in this case is $g = 1$. We note the equivalence between Riemann theta functions and Jacobian theta functions in the genus-1 case, however, working with the former allows for the generalization of this framework to the multimode or $g > 1$ case. 

We find the final expression for $\bar r^\mathrm{(cat)}_\mu(\boldsymbol{t})$ by replacing $\catgk$ in~\cref{eqn:cat_state_overlap_gk} with~\cref{eqn:cat_gk}, and then normalizing the Bloch vector coefficients with~\cref{eqn:r_overlap_normalization}. 

\begin{figure*}[!htbp]
\centering
\subfloat[]{\label{fig:cat_2legs}%
  \includegraphics[width=0.45\linewidth]{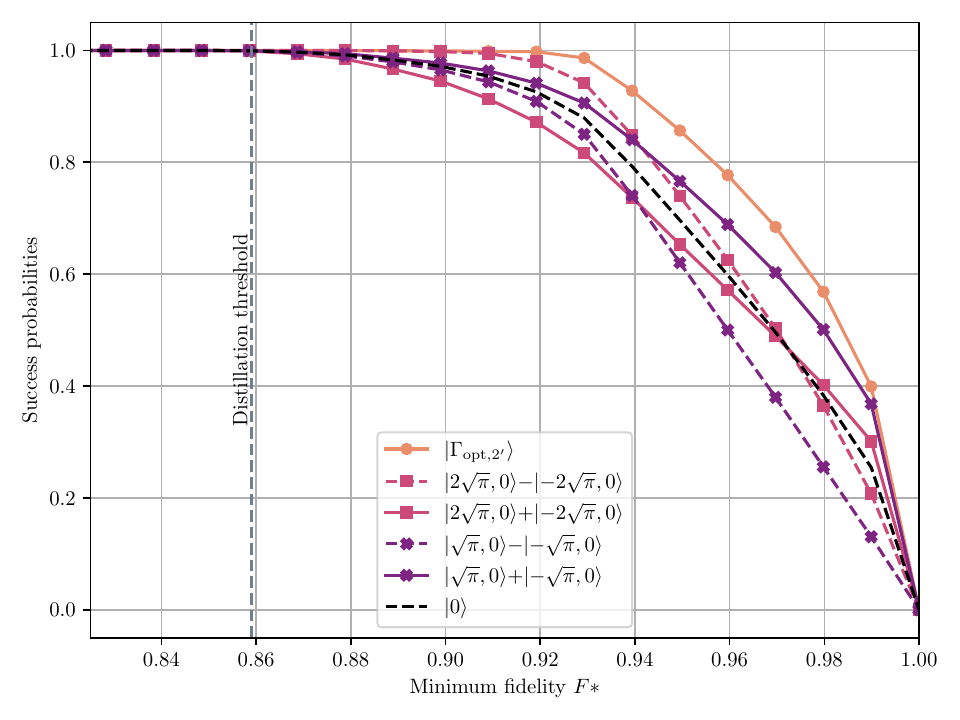}}%
\hspace{0.00\linewidth}%
\subfloat[]{\label{fig:cat_2legs_positions}%
  \includegraphics[width=0.33\linewidth]{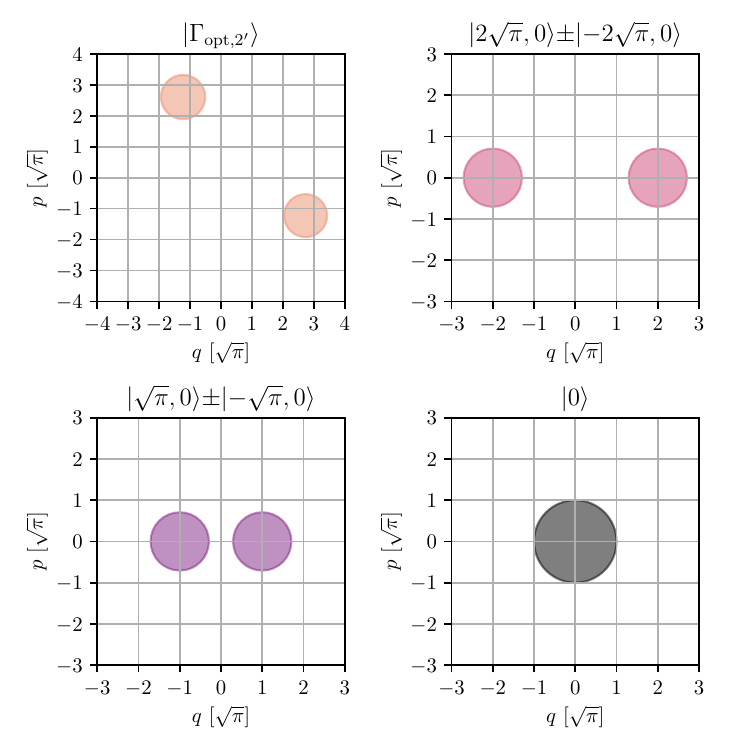}}%
\hspace{0.00\linewidth}%
\subfloat[]{\label{fig:cat_trunc_h_state}%
  \includegraphics[width=0.45\linewidth]{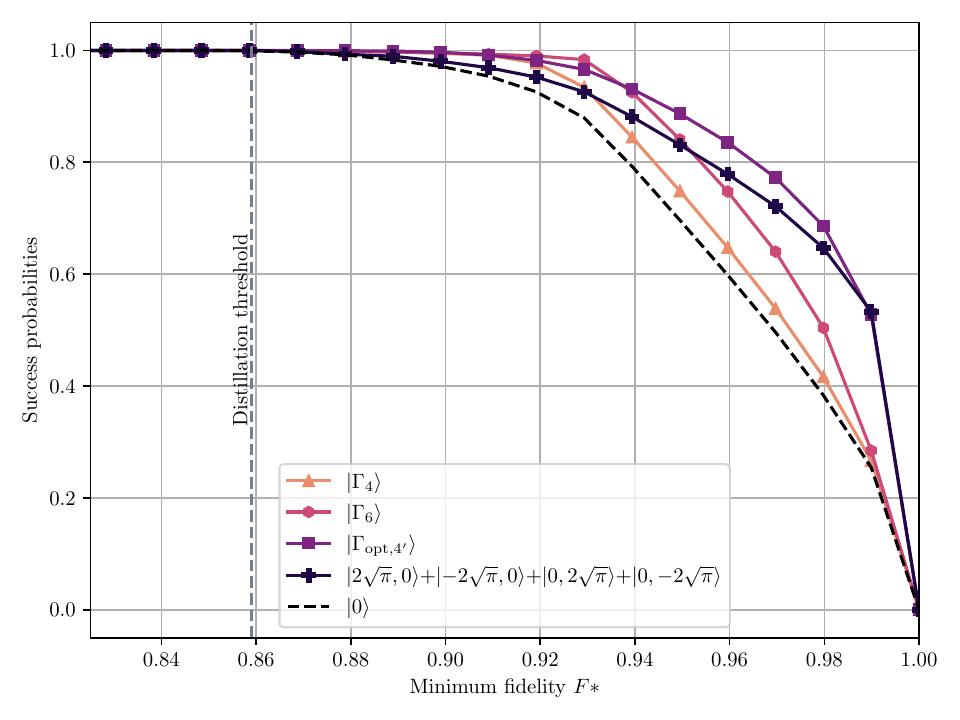}}%
\hspace{0.00\linewidth}%
\subfloat[]{\label{fig:cat_trunc_positions}%
  \includegraphics[width=0.33\linewidth]{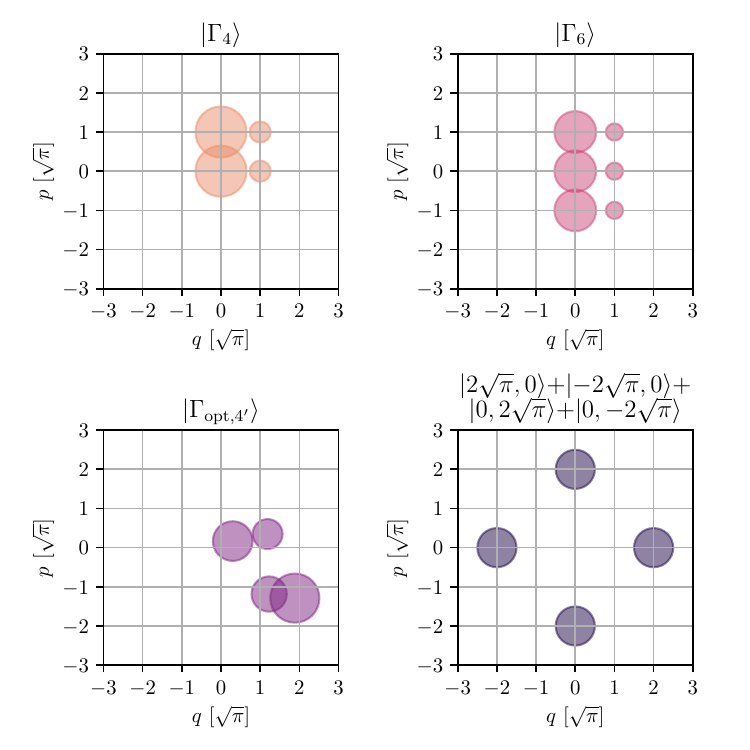}}%
\hspace{0.00\linewidth}%
  \subfloat[]{\label{fig:cat_states_opt_sprobs}%
  \includegraphics[width=0.45\linewidth]{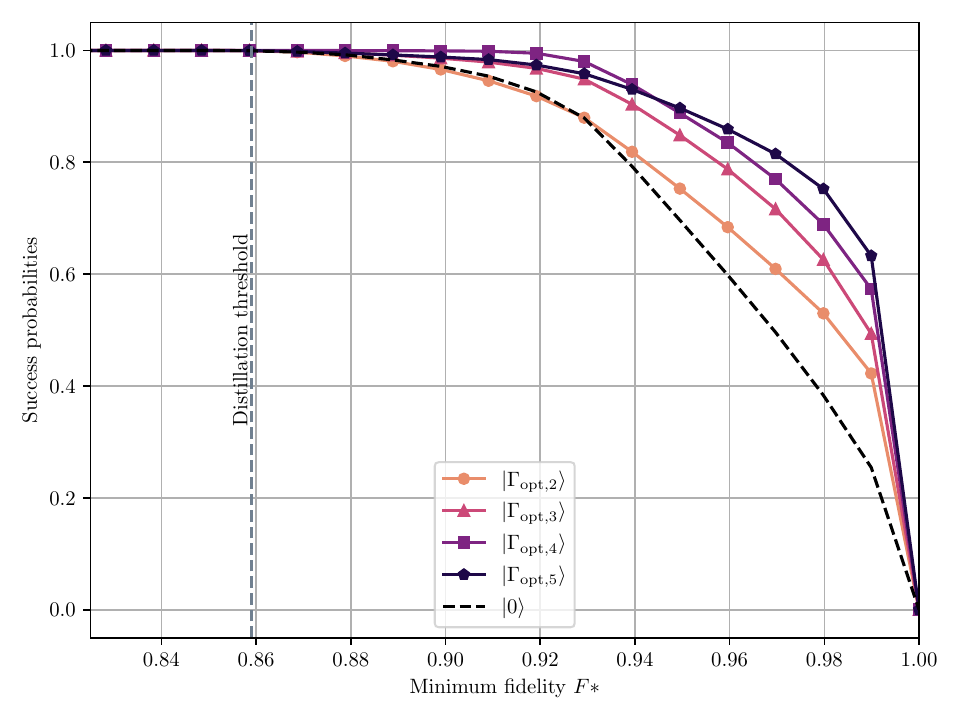}}%
\hspace{0.00\linewidth}%
\subfloat[]{\label{fig:cat_opt_positions}%
  \includegraphics[width=0.33\linewidth]{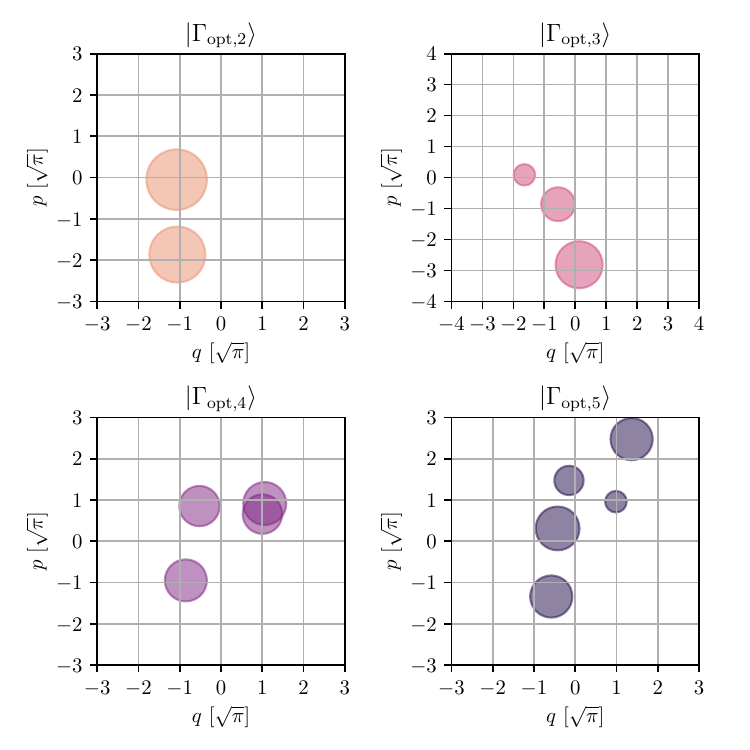}}%
\hspace{0.00\linewidth}%
\caption{These figures are to be read as follows, the graphs on the left contain the success probabilities of cat states as a function of its minimum fidelity $F^*$, the adjacent plots on the right are the respective positions of these cat states in phase space. (a) The success probabilities for two-legged cat states. Here we use the notation $\ket{q_0, p_0} \pm \ket{q_1, p_1}$ to indicate the positions of the coherent states in phase space. In general the even-symmetric cat states have higher success probabilities, i.e., they have higher chances of projecting close to an $H$-type state. However the $\ket{2\sqrt{\pi},0} - \ket{-2\sqrt{\pi},0}$ state on the GKP lattice has higher success probabilities than most odd cat states. It also performs better than its even counterpart for  $F^* \lesssim   0.97$, when they cross. The odd $\ket{\sqrt{\pi} , 0} - \ket{-\sqrt{\pi},0}$ state performs the worst out of all of these states despite being situated on the GKP grid. (b) Diagrammatic representation of the coherent states in phase space. The radii are functions of the coefficient of each state. These representations are not to be confused with Wigner functions.(c) The four-legged ($\ket{\Gamma_4}$) and six-legged ($\ket{\Gamma_6}$) truncated H states compared to a balanced four-legged cat state, and an optimized four-legged cat state. (d) Diagrammatic representation of the truncated $\ket{+H_L}$ state alongside the compass state, situated on the GKP grid, and an optimized cat state. (e) Optimized cat states and their success probabilities. All states were optimized at $F^* = 0.99$. (f) Diagrammatic representation of the optimized cat states.} 
\label{fig: cat_states}
\end{figure*}

\subsubsection{Numerical results}
The final expression above can be used to numerically compute the best fidelity for any input cat state and any given syndrome. In turn, this can be used to compute the success probability as a function of $F^*$, which is displayed in~\cref{fig: cat_states} for several input cat states. Here, the success probability is found as a function of the minimum fidelity $F^*$, where for each $F^*$, the success probability was calculated with~\cref{eqn:success_probability}, via a numerical integration method, enabled by a fast Riemann theta function evaluation modelled after the SageMath~\cite{sagemath} implementation using the Abelfunctions library~\cite{swierczewski_abelfunctions}. The integral is over the syndrome $\boldsymbol{t}\in \left[-\sqrt{\pi}, \sqrt{\pi}\right)^2$ space and is evaluated by discretizing the space in a grid of $400\times 400$ points for $t_q,\ t_p$. More details can be found in Appendix~\ref{app:numerics_details_sprobs}.

In order for a state to be distillable, it needs to pass the distillation threshold set by the specific distillation code. Here, we choose the 15-to-1 qubit Reed--Muller code as the baseline, and its distillation threshold is represented with a vertical line at $F^* = 0.859$, and we see that all input states can surpass this threshold with non-zero probability. Further, the vacuum state success probability curve, depicted by a dashed black line, represents the benchmark to which all input states are compared. Generally, a curve above this benchmark will indicate better performance.

Figure \ref{fig:cat_2legs} shows the success probabilities for various two-legged cat states. A state of the form $\ket{2\sqrt{\pi}, 0} + \ket{-2\sqrt{ \pi},0}$ ($\ket{2\sqrt{\pi}, 0} - \ket{-2\sqrt{ \pi},0}$) represents an even (odd) state whose coherent state components are situated at positions $q_0 = 2\sqrt{\pi}, p_0 = 0$ and $q_1 = -2\sqrt{\pi}, p_1 = 0$, respectively, with the normalisation omitted for brevity. These are pictorially represented in ~\cref{fig:cat_2legs_positions} \footnote{Note that this diagram does \textbf{not} represent the Wigner function of the state.}. In general, we find odd cat states to perform worse than both even cat states and the vacuum state. However, one notable exception is the $\ket{2\sqrt{\pi}, 0} - \ket{-2\sqrt{ \pi}, 0}$ state that outperforms its even counterpart for $F^* \lesssim 0.97$, where they cross.

Intuitively, one could assume that placing the coherent states on the GKP grid would lead to better performance. Indeed, one might start from the expansions~\cite{Grimsmo_2021}
\begin{align} \label{eqn: gkp_grimsmo_puri_alt}
    \ket{0_L} &\propto \sum_{k,l = -\infty}^{\infty} \hat{S}_x^k\bar{Z}^l  \ket{0} = \sum_{\alpha \in \mathcal{L}_1, \alpha' \in \mathcal{L}_2} D(\alpha)D(\alpha')\ket{0} \\
    \ket{1_L} &\propto \sum_{k,l = -\infty}^{\infty} \hat{S}_z^k\bar{Z}^l  \ket{0} = \sum_{\beta \in \mathcal{L}_3, \beta' \in \mathcal{L}_4} D(\beta)D(\beta') \ket{0},
\end{align} then write, using ~\cref{eqn:H_plus_state} \begin{equation}
    \ket{+H_L} = \cos\left(\frac{\pi}{8}\right)\ket{0_L} + \sin\left(\frac{\pi}{8}\right)\ket{1_L}
\end{equation} and truncate the resulting sum, retaining only the coherent states of smallest amplitude, and expect to find the optimal input state for a given number of components. However this assumption does not always hold. For instance, we see that the odd $\ket{\sqrt{\pi},0} - \ket{-\sqrt{\pi},0}$ state performs poorly compared to the benchmark state, despite being located on the GKP lattice points. 

To further investigate the role of symmetry, we optimized the coefficients and components of the cat state in~\cref{eqn:gen_cat_state} to maximize the success probability, for different numbers of legs. We used the SciPy differential evolution global optimizer, and we maximized two different objective functions --- the \textit{point} method (maximize the success probability at fixed $F^*$) and the \textit{area} method (maximize the area under the curve), see Appendix~\ref{app:numerics_details_optimization} for more details. For brevity, Fig.~\ref{fig: cat_states} only shows the curves obtained with the former.

We find several optimized cat states, and denote them as $\ket{\Gamma_\mathrm{opt,L}}$, where the subscripts `opt' and `L' denote that it is an $L$-legged optimized cat state. In ~\cref{fig:cat_2legs}, we find that $\ket{\Gamma_\mathrm{opt,2'}} \approx (0.679 - 0.412 i)\ket{2.734 \sqrt{\pi}, - 1.226 \sqrt{\pi}} + (0.801 + 0.159 i) \ket{- 1.226 \sqrt{\pi}, + 2.607 \sqrt{\pi}} $, performs the best out of all the states shown, and it is not on the GKP grid; it was optimized over $F^* = 0.96$. 
We find several more optimized cat states, and see that while success probabilities do improve with a larger superposition number, the improvement is also regime-dependent. We also find that, in general, coherent states in these optimized superpositions are not on the GKP lattice points, see ~\cref{fig:cat_opt_positions}. Figure~\cref{fig:cat_states_opt_sprobs} compares the two-legged, three-legged, four-legged and five-legged optimized cat states. See Appendix~\ref{appendix: opt_state_descriptions} for the state descriptions. 

Figure \ref{fig:cat_trunc_h_state}, shows the success probabilities for the truncated four-legged and six-legged $+H$ state, which are $\ket{\Gamma_4}$ and $\ket{\Gamma_6}$, respectively. We see that both these states outperform the vacuum state, however we also contrast them with a four-legged cat state $\ket{\Gamma_\mathrm{opt,4'}}$ optimized at $F^* = 0.95$, and see that it overtakes $\ket{\Gamma_6}$ in the $F^*  \gtrsim 0.94$ regime. We compare these curves to the balanced four-legged cat state on the GKP grid, $\ket{2\sqrt{\pi},0} + \ket{-2\sqrt{\pi},0} + \ket{0,2\sqrt{\pi}} + \ket{0,-2\sqrt{\pi}}$, and find that it outperforms the truncated states beyond the $F^* \gtrsim 0.95$ regime but matches the performance of the optimized cat state in the $F^* \gtrsim 0.99$ region. 

\subsection{Fock states}
In this section we perform a similar analysis for Fock states. All Fock states other than the vacuum state are non-Gaussian but, like cat states, they are routinely generated in experiments in a wide range of platforms~\cite{Cooper_2013, j7yp-fg9g}. We consider generic superpositions of finitely many Fock states 
\begin{equation}\label{eqn:fock_gen_superposition}
    \ket{\Phi_{n}} = \sum_{i=0}^n \phi_i\ket{i},
\end{equation}
where $\phi_i \in \mathbb{C}$.

\subsubsection{Bloch vector calculations \label{sec:bloch_vec_calcs_fock}}

We essentially follow the same procedure as in~\cref{subsubsec:cat_states_bloch_comps} to compute the Bloch vector components for error-corrected Fock states. Starting with~\cref{eqn:overlap}, the input state is $\hat\rho_\mathrm{in} = \hat \rho_n = \ketbra{n}$, and
\begin{align}
    \bar r^{(n)}_\mu(\boldsymbol{t}) & = \Tr[\hat{V}(-\boldsymbol{t})\ketbra{n}\hat{V}^\dagger(-\boldsymbol{t})\hat{\sigma}_L^\mu]\\
    & = \sum_{jk} \sigma_{jk}^\mu \bra{n}\hat{V}(\boldsymbol{t})\ket{j_L}\underbrace{\bra{k_L}\hat{V}(\boldsymbol{-t})\ket{n}}_{\fockgk}.
\end{align}
Once more, since $k \in \{0,1\}$ it is sufficient to find an expression for
\begin{align}
    \fockgk & =  \frac{1}{2 \sqrt[4]{\pi}} \sum_l e^{-i t_p (2l + k)\sqrt{\pi}}\underbrace{{}_q\bra{(2l + k)\sqrt{\pi} +t_q} \ket{n}}_{\text{Fock state wavefunction}}.\label{eqn:fock_1_gkp_wf}
\end{align}
 The wavefunction of the Fock state reads~\cite{ulf_leonhardt_text} 
\begin{equation} \label{eqn:wf_fock_state}
    \psi_n(q) = \frac{1}{\sqrt{2^n n!}} \frac{1}{\pi^{1/4}} e^{-\frac{q^2}{2}} H_n(q),
\end{equation}
where $H_n(q)$ is the physicist's Hermite polynomial
\begin{equation}\label{eqn:hermite_polynomial}
    H_n(q) = (-1)^n e^{q^2}\frac{d^n}{dq^n}e^{-q^2}.
\end{equation}
As an example, we show here the detailed derivation for $\ket{1}$, those for $n$ up to four are shown in Appendix~\ref{app:fock_states_rmus}. The Hermite polynomial for this state is $H_1(q) = 2q$, then the wavefunction is defined as
\begin{equation} \label{eqn:wf_fock_1}
    \psi_1(q) =  \frac{1}{\pi^{1/4}\sqrt{2}} e^{-\frac{q^2}{2}} 2q.
\end{equation}
The wavefunction defined in ~\cref{eqn:fock_1_gkp_wf} becomes
\begin{equation}\label{eqn:fock_gkp_wavefunction}
\begin{split}
    & {}_q\bra{(2l + k)\sqrt{\pi} +t_q} \ket{n} \\
    & = \frac{ e^{\left(-2l^2 \pi - 2lk\pi - \frac{k^2 \pi}{2} - 2l\sqrt{\pi}t_q - k\sqrt{\pi}t_q - \frac{t_q^2}{2}\right)}(4l\sqrt{\pi} + 2k\sqrt{\pi} + 2t_q)}{\sqrt{2}\pi^{1/4}}.\\
\end{split}
\end{equation}
Substituting ~\cref{eqn:fock_gkp_wavefunction} into ~\cref{eqn:fock_1_gkp_wf} with $T = t_q + it_p$ provides the expression for the overlap
\begin{widetext}
\begin{equation}
\begin{split}
    \fockgk[1] & = \frac{1}{2 \sqrt[4]{\pi}}\underbrace{\frac{\exp\left[ - \frac{k^2 \pi}{2} - k\sqrt{\pi} T - \frac{(\Re{T})^2}{2} \right]}{\sqrt{2}\pi^{1/4}}}_{\text{Gaussian term  } G_1(\boldsymbol{t})}\times \\ & \left(\underbrace{ \sum_l (4 l \sqrt{\pi}) \exp\left [2\pi i \left ( \frac{1}{2} \underbrace{(2i)}_{\Omega_{\bar n}} l^2 + l\underbrace{\left( ik + \frac{iT}{\sqrt{\pi}}\right)}_{z_{\bar n}(\boldsymbol{t})}\right) \right]}_{\text{first derivative of the Riemann theta function }\frac{d\Theta(z_{\bar n}(\boldsymbol{t}), \Omega_{\bar n})}{dz_{\bar n}}} + (2k\sqrt{\pi} + 2t_q)
   \underbrace{\sum_l\exp\left [2\pi i \left ( \frac{1}{2} \underbrace{(2i)}_{\Omega_{\bar n}} l^2 + l\underbrace{\left( ik + \frac{i T}{\sqrt{\pi}}\right)}_{z_{\bar n}(\boldsymbol{t})}\right) \right]}_{\text{Riemann theta function }\Theta(z_{\bar n}(\boldsymbol{t}), \Omega_{\bar n})}\right).
\end{split}
\end{equation}
\end{widetext}
\noindent This expression includes a first derivative of the Riemann theta term. Indeed, the derivative  with respect to $z_{\bar n}$ reads
\begin{equation}\begin{aligned}
    \frac{d\Theta(z_{\bar n}(\boldsymbol{t}), \Omega_{\bar n})}{dz_{\bar n}}& =\\&\hspace{-0.5cm} \sum_l (2 \pi i l) \exp\left [2\pi i \left ( \frac{1}{2} \Omega_{\bar n} l^2 + z_{\bar n}(\boldsymbol{t}) l \right )\right].
\end{aligned}
\end{equation}
Finally, we obtain:
\begin{equation}\begin{aligned}
    \bra{k_L}\hat{V}(-\boldsymbol{t})\ket{1}& = \frac{1}{2 \sqrt[4]{\pi}}\frac{1}{\sqrt{2}\pi^{1/4}} G_{\bar 1}(\boldsymbol{t}) \\ &\times\left(A_1 \frac{d\Theta(z_{\bar n}(\boldsymbol{t}), \Omega_{\bar n})}{dz_{\bar n}} + A_0 \Theta(z_{\bar n}(\boldsymbol{t}), \Omega_{\bar n})\right),
\end{aligned}
\end{equation}
where $G_{\bar1}(\boldsymbol{t})$ represents the Gaussian term for $\ket{1}$, $A_1 = \frac{-2 i}{\sqrt{\pi}}$, $A_0 = (2k \sqrt{\pi} + 2t_q)$, and $\Theta(z_{\bar n}(\boldsymbol{t}), \Omega_{\bar n})$ is the Riemann theta function defined in~\cref{eqn:riemann-theta} with 
\begin{equation}\label{eqn:riemann-theta-params-fock}
    z = z_{\bar n}(\boldsymbol{t}) = ik - \frac{t_p}{\sqrt{\pi}} + \frac{i t_q}{\sqrt{\pi}} \text{ and } \Omega = \Omega_{\bar n} = 2i
\end{equation}
The overlaps for the GKP error-corrected Fock states are similar to those of cat states with additional derivatives of Riemann theta functions.
In general we have
\begin{equation}
    \fockgk = \frac{1}{2 \sqrt[4]{\pi}} G_{\bar n}(\boldsymbol{t}) \left ( \sum_{i=0}^n A_i \frac{d^i\Theta( z_{\bar n} ,\Omega_{\bar n} )}{dz_{\bar n}^n} \right),
\end{equation}
 where the arguments of the Riemann theta term are as in ~\cref{eqn:riemann-theta-params-fock} and the Gaussian term $G_n(\boldsymbol{t})$ is 
\begin{equation}
    G_{\bar n}(\boldsymbol{t}) = \frac{1}{\pi^{1/4} \sqrt{2^n n!}}\exp\left[-\left(\frac{k \sqrt{\pi} + t_q}{\sqrt{2}}\right)^2 - i k\sqrt{\pi}t_p \right].
\end{equation}
The analytical expressions of $\fockgk$ for $n \in \{0,2,3,4\}$ can be found in Appendix~\ref{app:fock_states_rmus}.

The overlap with $\hat{\sigma}^1_L$ can be found using Eqs.~\eqref{eqn:gkp_logical_paulis_sigma_mu} and~\eqref{eqn:overlap}
\begin{equation}
\begin{split}
    \bar r_1^{(n)} (\boldsymbol{t}) &= \underbrace{\bra{n}V^\dagger(-\boldsymbol{t})\ket{0_L}}_{{f_0^{(n)}(\boldsymbol{t})}^*} \underbrace{\bra{1_L}V(-\boldsymbol{t})\ket{n}}_{f_1^{(n)}(\boldsymbol{t})} \\
    & + \underbrace{\bra{n}V^\dagger(-\boldsymbol{t})\ket{1_L}}_{{f_1^{(n)}(\boldsymbol{t})}^*}  \underbrace{\bra{0_L}V(-\boldsymbol{t})\ket{n}}_{f_0^{(n)}(\boldsymbol{t})},\\
\end{split}
\end{equation}
The same procedure can be applied to find the overlaps with $\hat{\sigma}^0_L, \hat{\sigma}^2_L$ and $\hat{\sigma}^3_L$. These Bloch vector coefficients can then be normalized as $r_\mu^{(n)} (\boldsymbol{t}) = \frac{\bar{r}_\mu^{(n)} (\boldsymbol{t})}{\bar{r}_0 (\boldsymbol{t})}$. 

\begin{figure}[t]

\includegraphics[width=1\linewidth]{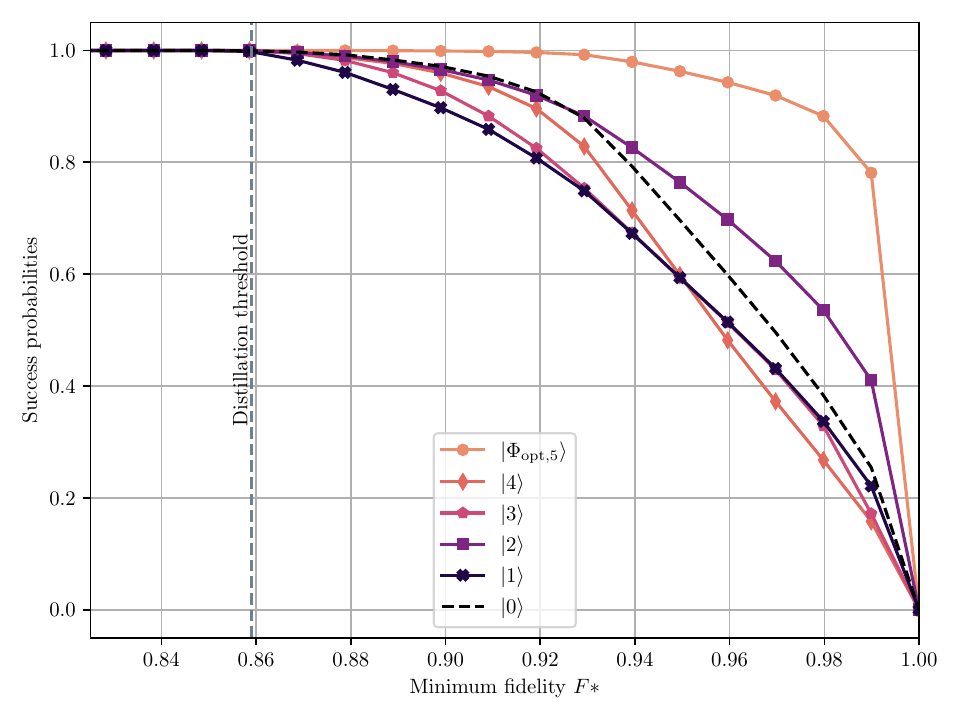}
\caption{The success probability curves for Fock states. The Reed--Muller distillation threshold of $F^* > 0.859$, is maintained for the protocol with Fock states. The best success probability is achieved by the optimized superposition state $\ket{\Phi_{\mathrm{opt},5}}$. Not only does $\ket{2}$ perform the best among the number states, it also outperforms the vacuum state. The odd Fock states $\ket{1}$ and $\ket{3}$ have success probabilities lower than $\ket{0}$. Notably, the even-symmetric $\ket{4}$ performs worse than the vacuum state, and worse than the odd-symmetric $\ket{1}$ in certain regions.}
\label{fig:fock_states_sprobs}
\end{figure}

\subsubsection{Optimized results}

As in the case of cat states, we evaluate the performance of Fock states $\ket{0},\ldots, \ket{4}$, alongside an optimized superposition of the five states, by generating their success probabilities, see~\cref{fig:fock_states_sprobs} for the results. 

All states have fidelities higher than the Reed--Muller distillation threshold of $F^* > 0.859$ with probabilities close to one. Of the number states, $\ket{2}$ performs the best, i.e., it has the highest success probability across a wide range of $F^*$, and its fidelities are higher than the vacuum state benchmark. In the case of cat states, a positive relationship between success probabilities and even-symmetry was observed, for most states, with notable exceptions. Further, in most regimes, a larger number of superposition coherent states presented higher success probability values. For Fock states, a similar trend could be expected. However, we note that $\ket{4}$ performs significantly worse compared to the vacuum state, $\ket{2}$, and in regimes of $F^* > 0.94$, even worse than odd-symmetric states. One pattern that is consistent with cat states is that the odd-parity $\ket{1}$ and $\ket{3}$ perform worse than vacuum. 

We optimized over the amplitudes $\phi_i$ of the superposition in~\cref{eqn:fock_gen_superposition} to maximize the success probability at $F^* = 0.98$. The best state we found is $\ket{\Phi_\mathrm{opt,5}} \approx (0.899-0.716i)\ket{0} +(-0.644+0.451i) \ket{1} + (-0.104-0.167i)\ket{2} +(-0.638+0.037i) \ket{3} +(-0.895+0.599i) \ket{4}$. Here the subscripts `opt' and `5' represent a state optimized over the first 5 Fock states (up to $n=4$). 
\section{Practical Advantages} \label{sec: practical_advntgs}
The previous sections established that GKP error-correcting cat states and Fock states can produce distillable magic states with higher success probabilities than the vacuum state. In this section we ask whether such improvements can lead to practical advantages when the overhead of the subsequent distillation is taken into account. Namely, we estimate how many copies of each input state \emph{before} error correction are required to achieve a given fidelity $F_\mathrm{target}$ with a target magic state (class) \emph{after} distillation. We will refer to this as \textit{distillation cost}. We demonstrate that higher success probabilities do lead to lower distillation cost.

\subsection{Estimate of distillation cost}
As mentioned in~\cref{subsec: msd}, many different distillation schemes have been proposed and studied in the literature. Here we take the protocol based on the 15-to-1 qubit Reed--Muller code as reference~\cite{Bravyi_2005}. Recall that this procedure consumes $15$ noisy magic states to distill $1$ magic state closer to the target class; this process can be iterated recursively until the output qubit has fidelity higher than a desired target, $F_\mathrm{target}$. In the case of GKP logical qubits, i.e., encoded qubits, one needs to consider the additional CV resource states required to create each distillable logical magic state, since the success probability is lower than one. For example, if one requires $F^*\geq .96$ to start the distillation protocol, it will take on average 25 attempts at error-correcting vacuum states to generate the 15 copies of encoded magic states needed for one distillation step, since the success probability is around $60\%$ (see Fig.~\ref{fig:fock_states_sprobs}).

We introduce $\nu$, the number of qubit states required for the entire distillation procedure when working in the Reed--Muller configuration. This number depends on the input state fidelity, $F^* = 1-\epsilon$, and the target fidelity, $F_\mathrm{target} = 1 - \epsilon_\mathrm{target}$, and was estimated in~\cite{Bravyi_2005} to be
\begin{equation} \label{eqn:nu_qubit_number}
    \nu (\epsilon, \epsilon_\mathrm{target}) = \prod_{i = 1}^{\textrm{rounds}} 15 \times\frac{16}{1 + 15 (1 - 2 \epsilon_{i -1})^8},
\end{equation}
where 
\begin{equation}
    \epsilon_i = \frac{1 - 15(1-2\epsilon_{i-1})^7 + 15(1-2\epsilon_{i-1})^8 - (1-2\epsilon_{i-1})^{15}}{2\left(1 + 15(1-2\epsilon_{i-1})^8\right)}
\end{equation}
for $ \epsilon_{0} = \epsilon$. This recursive process continues until $\epsilon_i \leq \epsilon_\mathrm{target}$. Here the focus is on the 15-to-1 Reed--Muller code for illustrative purposes, however, we expect similar conclusions to hold for other choices of distillation codes such as the 7-to-1 Steane code~\cite{Reichardt_2005}.

In the previous sections, we computed the probability of successfully obtaining a distillable GKP logical magic state with a given fidelity $F^*$ through error correction of an input state. The average number of required CV input states for the distillation procedure is related to the success probabilities that have been obtained for the same $F^*=1-\epsilon$. The estimated CV distillation cost is then
\begin{equation}\label{eqn:N_cv_resources}
    N_{\mathrm{CV}_\mathrm{state}}(\epsilon, \epsilon_\mathrm{target}) = \frac{\nu(\epsilon, \epsilon_\mathrm{target})}{\Pr[F^* \geq 1 - \epsilon]},
\end{equation}
where $\Pr[F^* \geq 1 - \epsilon]$ is the success probability at a given fidelity $F^*$.

\begin{figure}[b]
    \centering
    \includegraphics[scale=0.53]{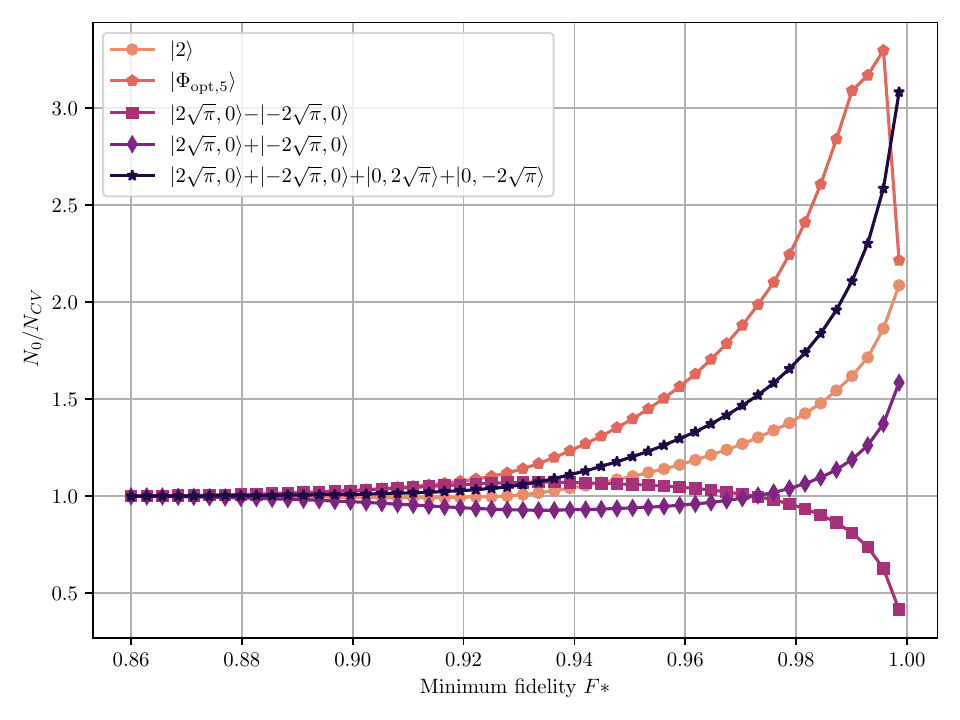}
    \caption{Ratio of the vacuum resources to the new non-Gaussian input states for $F_\mathrm{target} = 0.99$. The vacuum state consumes twice or thrice as many resources as $\ket{2}$, $\ket{\Phi_\mathrm{opt,5}}$ and $\ket{2\sqrt{\pi}, 0} + \ket{-2\sqrt{\pi}, 0} + \ket{0,2\sqrt{\pi}} +\ket{0,-2\sqrt{\pi}}$.}
    \label{fig:N_resources_cv_ratio}
\end{figure}

\begin{figure}[t]
    \centering
    \includegraphics[scale=0.53]{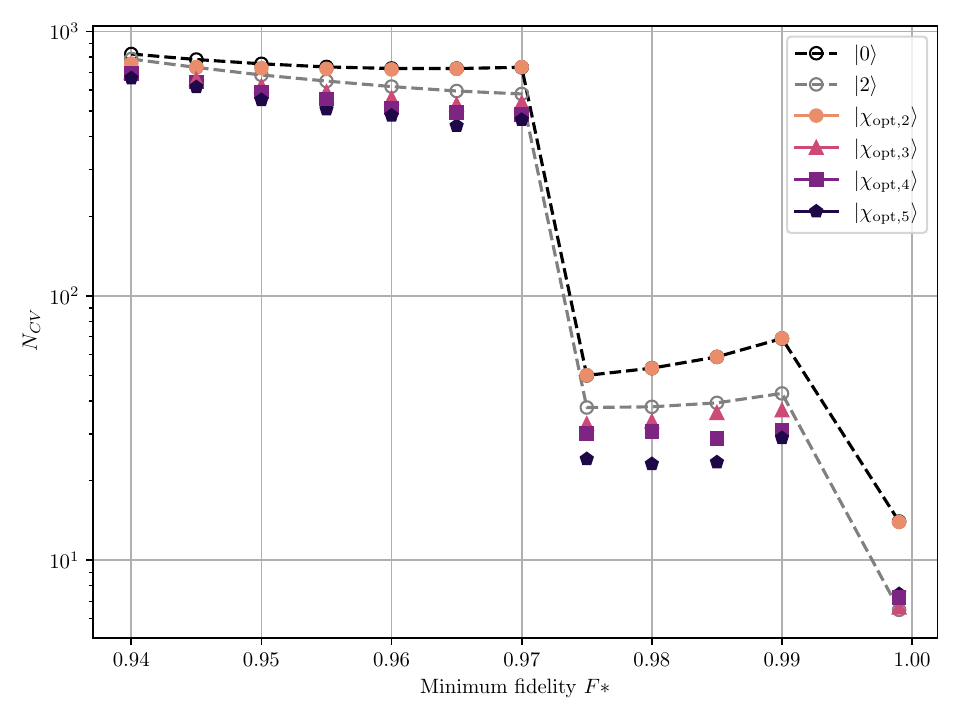}
    \caption{Log plot of $N_\mathrm{CV}$ for different superpositions of Fock states. The superposition weights were found by minimizing over the number of CV resources $N_\mathrm{CV}$ for different minimum fidelities $F^*$. The target fidelity of the output state was set at $F_\mathrm{target} = 0.999$. Here $\chi_\mathrm{opt,2} = \phi_0\ket{0} + \phi_1\ket{1}$, and $\chi_\mathrm{opt,3} = \phi_0\ket{0} + \phi_1\ket{1} + \phi_2\ket{2}$, and so on. We assume perfect syndrome extraction; therefore, experimental values are expected to differ.}
    \label{fig:N_resources_fock_per_fid}
\end{figure}

\subsubsection{Numerical results}
For $F_\mathrm{target} = 0.99$, we computed the distillation cost for various input cat and Fock states and compared it to that of the vacuum state. The ratio $N_{\mathrm{CV, \ket{0}}}/N_{\mathrm{CV, state}}$ is plotted in ~\cref{fig:N_resources_cv_ratio}. Relative to using vacuum states, $\ket{2}$ and the superposition state $\ket{\Phi_{\mathrm{opt},5}}$ allow one to reduce the (average) number of error correction rounds by factors of two and three respectively. For cat states, the $N_{\mathrm{CV, cat}}$ varies according to the regime. The $\ket{2\sqrt{\pi},0 } + \ket{-2\sqrt{\pi},0}$ state shows a maximum ratio of $\approx 1.5$ in the high fidelity regime but underperforms in the $F^* \lesssim 0.98$ region. However the $\ket{2\sqrt{\pi},0 } - \ket{-2\sqrt{\pi},0}$ does not show marked reduction in the number of resources. The four-legged compass state $\ket{2\sqrt{\pi},0 } + \ket{-2\sqrt{\pi},0} + \ket{0,2\sqrt{\pi} } + \ket{0,-2\sqrt{\pi}}$ shows large reductions in the $F^* \gtrsim 0.98$ region.

Many of these curves exhibit regime dependence. The $\ket{\Phi_{\mathrm{opt},5}}$ curve shows a steep and marked reduction in the resource cost until $F^* = 0.99$, when there is a sudden drop, however the reduction in the resources is still significant enough that this state remains a suitable candidate for distillation for high fidelity input states. 

To interpret these results, it is also important to consider the distillation cost itself, independently of that of the vacuum. We optimize the quantity $N_{\mathrm{CV}}$ for superpositions of up to five Fock states. For each minimum fidelity $F^*$ (after error correction), we minimize the distillation cost at a fixed output fidelity $F_\mathrm{target}$ after distillation. This optimization is feasible since the overlaps $f_k^{(n)}\left(\boldsymbol{t}\right)$ can be pre-computed for all values of $\boldsymbol{t}$ used for the numerical integration. Only the amplitudes $\phi_n$ in~\cref{eqn:fock_gen_superposition} are then optimization variables. For coherent states, the optimization should run over the displacements $\alpha_c$ as well, which means that theta functions would need to be computed at each evaluation of the objective function, making the optimization considerably slower. See Appendix~\ref{app:numerics_details_optimization} for more details on the implementation. The optimization results are found in~\cref{fig:N_resources_fock_per_fid} for $F_\mathrm{target} = 0.999$.

We see that the distillation cost decreases in steps with respect to the minimum fidelity, and we also observe that an increase in the number of superposition states decreases the distillation cost. The first regime is identified as that between $F^* = 0.94$ to $F^* = 0.97$, where $\ket{\chi_\mathrm{opt, 4}}$ requires less resources than $\ket{\chi_\mathrm{opt, 3}}$ and so on. The observed jumps are distinctive of the iterative nature of the distillation code itself. Moreover, we note that these curves must eventually spike as $F^* \to 1$, as a consequence of the vanishing success probabilities in the denominator of~\cref{eqn:N_cv_resources}.

A useful takeaway for applications is that it is better to post-select on higher-fidelity magic states after error correction, and run fewer distillation rounds. The obvious caveat is that a careful analysis in the presence of finite squeezing (non-ideal error correction) and other imperfections would be needed to assess the robustness of these results in a realistic scenario. 

\section{Bloch sphere probability density} \label{sec: output_distribs}

Thus far, we have established that certain cat states and certain Fock states do not produce distillable magic states of a higher fidelity than those obtained with the Gaussian vacuum state. When these states are error-corrected with the GKP code, a resulting change of structure in phase space takes place and these states get mapped to discrete logical GKP states. In this section, we investigate the origin of the high fidelity results of certain non-Gaussian states, and the low fidelity results of other non-Gaussian states. 

This is done by identifying the probability densities of the GKP logical output states on the Bloch sphere. Recall that each output state $\rho_{\mathrm{out}}(\boldsymbol{t})$ is described by a normalized Bloch vector which provides the coordinates for a single point on the surface of the Bloch sphere. We determined success probability by integrating over all possible syndrome $\boldsymbol{t}$ values in a given window, for each input state. Here we produced histograms displaying the probability of various states being projected into different zones of the logical Bloch sphere. This allows us to investigate the source of these observed success probability curves, to determine whether there is a bias (or lack thereof) in the projection process. In particular, we ask whether states leading to good performance are projected close to magic states, or whether the low-fidelity states do not get projected close to magic states. 

\begin{figure*}[!htbp]
\centering
\begin{tikzpicture}[
    every node/.style={inner sep=0, outer sep=0},
    panel/.style={anchor=north west},
]

\newlength{\panelw}\setlength{\panelw}{5.8cm}  
\newlength{\panelh}\setlength{\panelh}{5.0cm}  
\newlength{\hgap}\setlength{\hgap}{0.1cm}      
\newlength{\vgap}\setlength{\vgap}{0.4cm}      

\node[panel, label=below:{(a)}] (a) at (0, 0)
  {\includegraphics[width=\panelw]{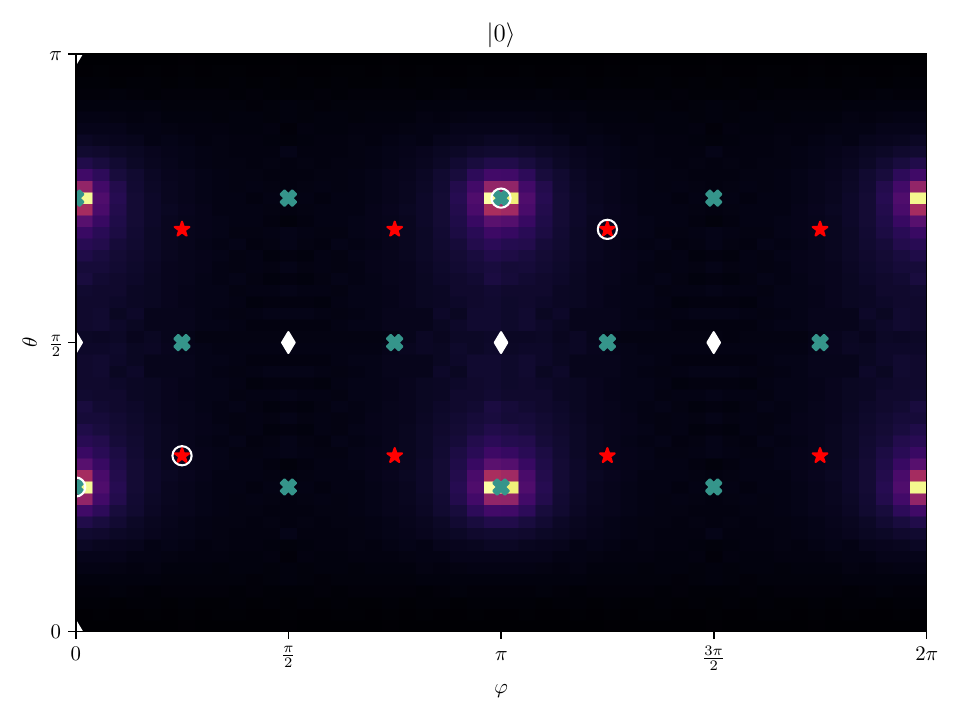}};
  \label{fig:fock_0_hist}
\node[panel, label=below:{(b)}] (b) at ($(a.north east) + (\hgap, 0)$)
  {\includegraphics[width=\panelw]{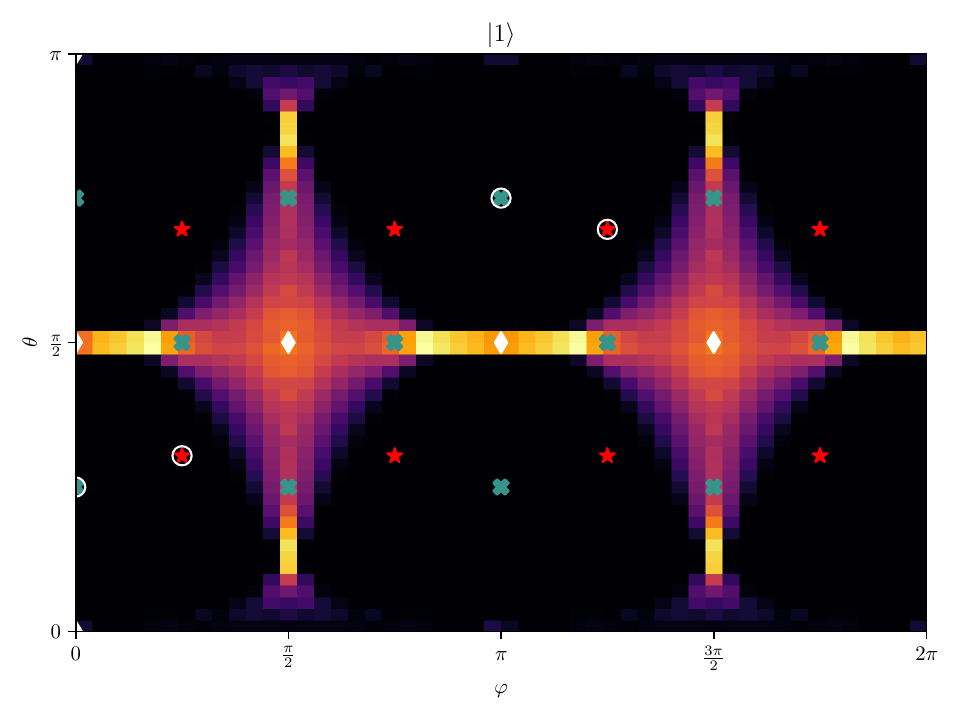}};
  \label{fig:fock_1_hist}
\node[panel, label=below:{(c)}] (c) at ($(b.north east) + (\hgap, 0)$)
  {\includegraphics[width=\panelw]{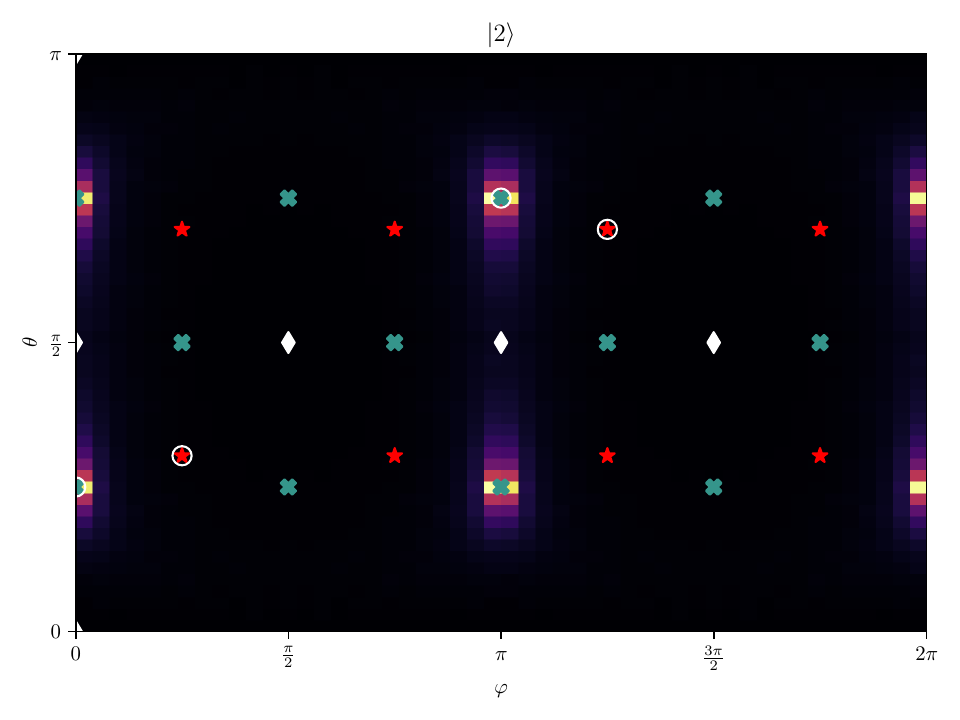}};
  \label{fig:fock_2_hist}

\node[panel, label=below:{(d)}] (d) at ($(a.south west) - (0, \vgap)$)
  {\includegraphics[width=\panelw]{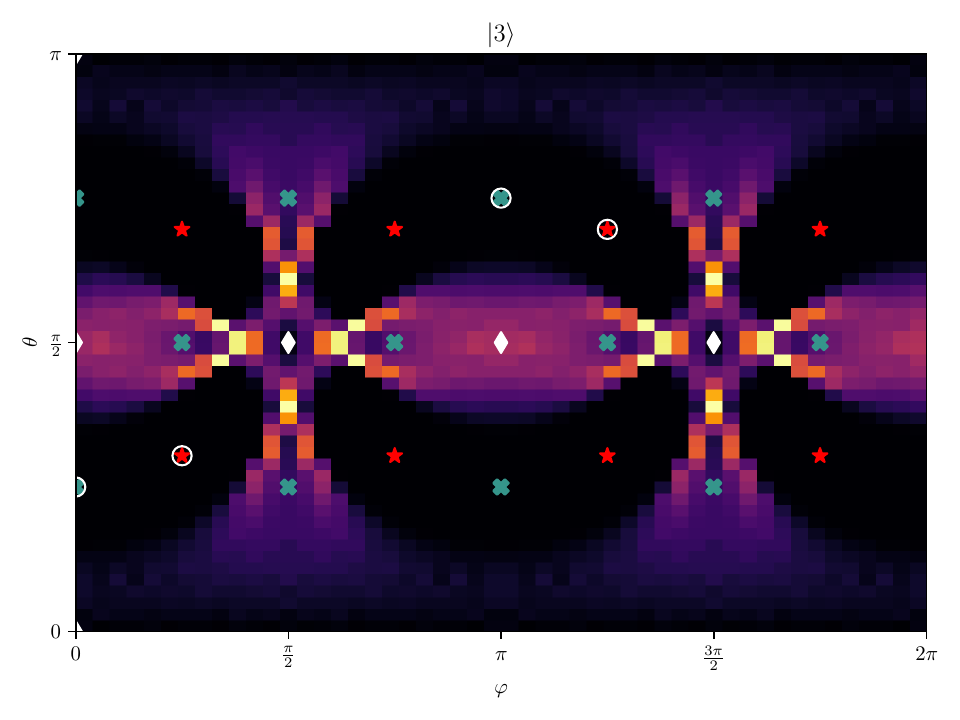}};
  \label{fig:fock_3_hist}
\node[panel, label=below:{(e)}] (e) at ($(d.north east) + (\hgap, 0)$)
  {\includegraphics[width=\panelw]{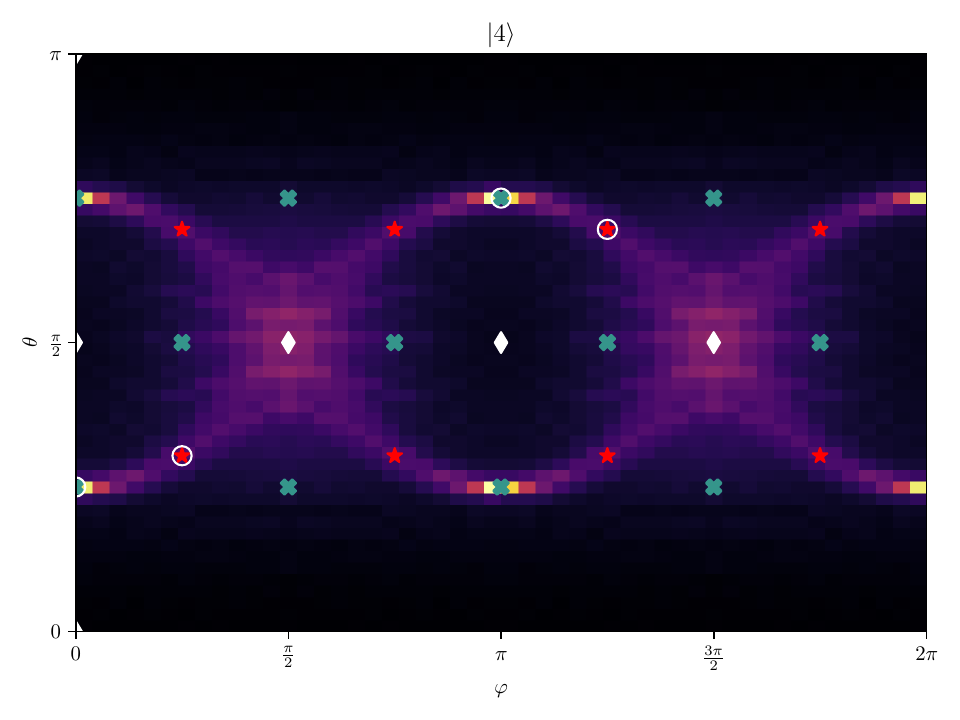}};
  \label{fig:fock_4_hist}
\node[panel, label=below:{(f)}] (f) at ($(e.north east) + (\hgap, 0)$)
  {\includegraphics[width=\panelw]{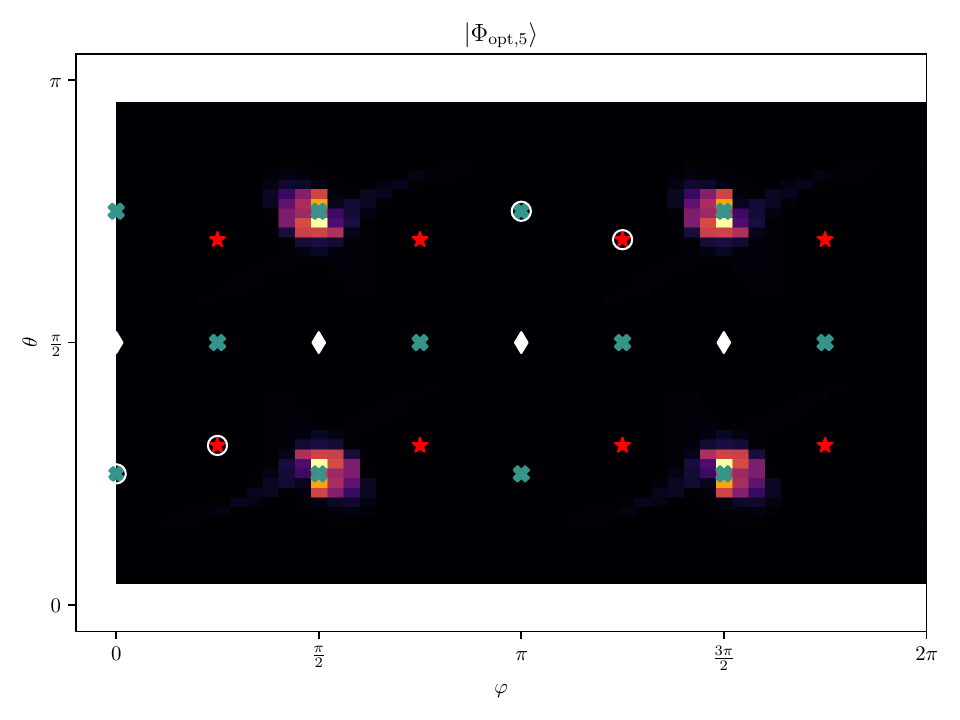}};
  \label{fig:fock_super_hist}

\pgfmathsetmacro{\gridwidth}{3*5.8+2*0.1}  
\pgfmathsetmacro{\rowgw}{2*5.8+0.3}        
\pgfmathsetmacro{\xoffset}{(\gridwidth-\rowgw)/2}  

\node[panel, label=below:{(g)}] (g) at ($(d.south west) - (0, \vgap) + (\xoffset cm, 0)$)
  {\includegraphics[width=\panelw]{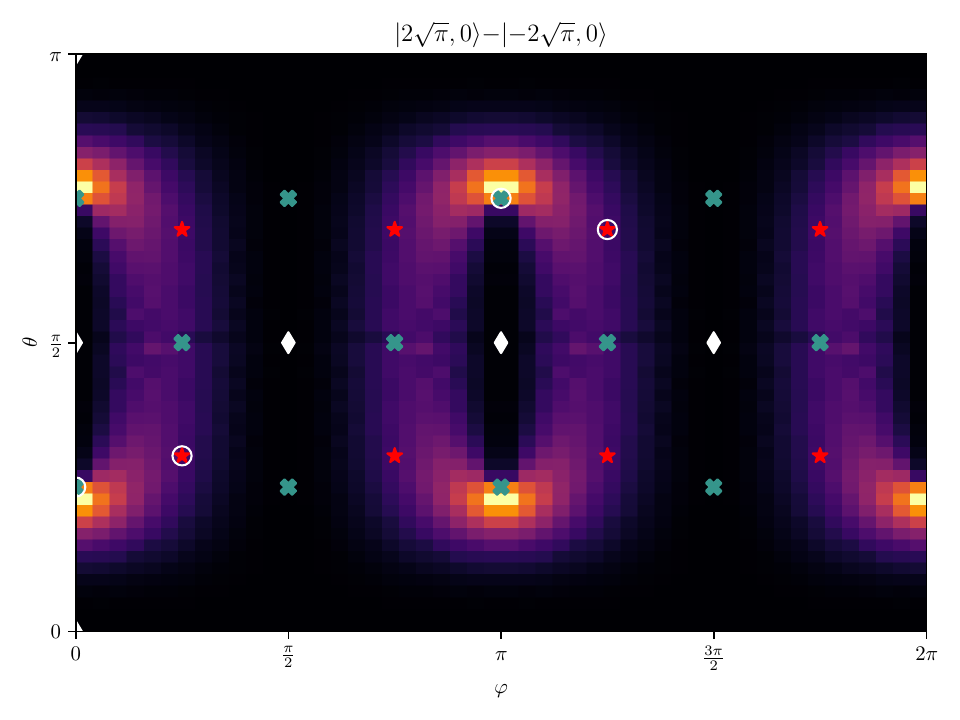}};
  \label{fig:cat_odd_2pi_hist}
\node[panel, label=below:{(h)}] (h) at ($(g.north east) + (\hgap, 0)$)
  {\includegraphics[width=\panelw]{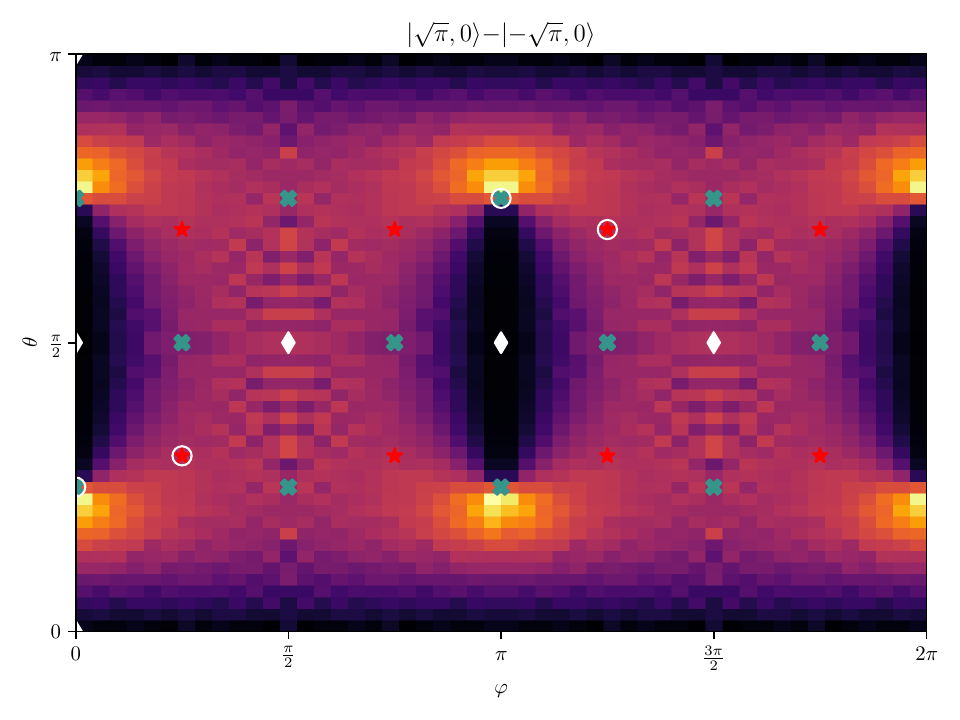}};
  \label{fig:cat_odd_pi_hist}

\node[anchor=north west] (cbar) at ($(c.north east) + (0.3cm, 0.0cm)$)
  {\includegraphics[height=\dimexpr2.5\panelh+\vgap\relax,
                     keepaspectratio]{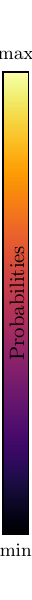}};

\node[anchor=north west] (legend) at ($(h.north east) + (0.3cm, -1.7cm)$)
  {\includegraphics[width=\dimexpr0.5\panelw\relax]{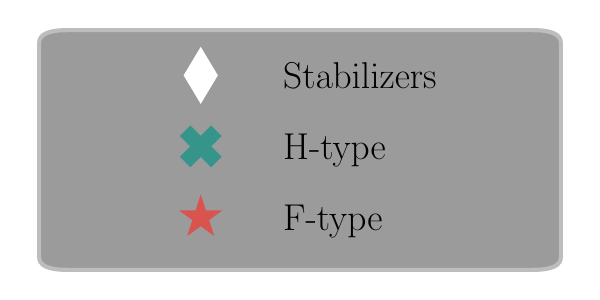}};

\end{tikzpicture}

\caption{Each Bloch vector, $\left(r_1(\boldsymbol{t}), r_2(\boldsymbol{t}), r_3(\boldsymbol{t})\right)$, is mapped to the spherical coordinates $(\theta, \varphi)$, that then serve as Cartesian coordinates for a 2D plane. The azimuthal angle $\varphi$ is depicted on the horizontal axis and the polar angle $\theta$ on the vertical axis. Colors correspond to the probability of the state being projected in the corresponding region. (a) This is the output state distribution of the GKP error-corrected vacuum state. The circled states are the $\pm$ $H$-type and $\pm$ $F$-type states. The color bar on the right indicates the sum of probabilities for a specific occurrence of a bin. States project close to the $H$-type magic states in the $X-Z$ plane with high probabilities. (b) From the success probability curves, it is known that $\ket{1}$ performs worse than the vacuum state. The output state distributions show that this state projects close to the Pauli eigenstates. It also shows low probability projection onto $H$-type states not in the $X-Z$ plane. (c)$\ket{2}$'s output distribution is similar to the vacuum state, in fact, all the even-symmetric states that surpass the vacuum benchmark have an output distribution that resembles this plot, where the states project onto the $H$-type magic states in the $X-Z$ plane. (d) $\ket{3}$ shows states projecting near the Pauli eigenstates with high probabilities, with low probability projection onto the $H$-type states outside the $X-Z$ plane. (e) $\ket{4}$ is a Fock state with even parity, it projects onto the $H$-type magic states with higher probabilities than it does onto the Pauli eigenstates, however its success probabilities curves are much lower than most other Fock states. There is also some notable projection onto the $F$-type states. (f) This is the output state distribution for an optimized Fock state that is in a unique superposition. The states project onto the $H$-type magic states in the $X-Y$ and $Y-Z$ plane, however they completely avoid projecting near the Pauli eigenstates. (g) An odd cat state whose success probabilities are higher than those of vacuum is the $\ket{2\sqrt\pi,0} - \ket{-2\sqrt\pi,0}$ state that lies on the GKP lattice. It has high probability projections onto the $H$-type magic states, moreover, it strays from the stabilizer states. (h) The state $\ket{\sqrt\pi,0} - \ket{-\sqrt\pi,0}$ also lies on the GKP grid, yet it performs exceptionally poorly when compared to other states. These projections show that the majority of states project directly onto the Pauli eigenstates and even though there exist high probability projections near the magic states, it is effectively cancelled out by the stabilizer projections.}\label{fig:histograms}
\end{figure*}

For visualization, histograms are displayed in Fig.~\ref{fig:histograms}, where the Bloch sphere is represented via a forward equirectangular projection~\cite{snyder1987}. Each Bloch vector, $\left(r_1(\boldsymbol{t}), r_2(\boldsymbol{t}), r_3(\boldsymbol{t})\right)$, is mapped to the spherical coordinates $(\theta, \varphi)$, that then serve as Cartesian coordinates for a 2D plane. We consider a $400\times400$ grid of $\boldsymbol{t}$ values in syndrome space, and compute the corresponding Bloch vectors. The Bloch sphere is then divided into ``squares'', or bins, where the color of each bin is determined by the sum of the probabilities of the Bloch vectors that are contained in it. 
The conversion from Bloch vectors to spherical coordinates requires the following equations
\begin{equation}
    \theta = \cos^{-1}{\left(r_3(\boldsymbol{t})\right)}
\end{equation}
\begin{equation}
    \varphi = \tan^{-1}\left({\frac{r_2(\boldsymbol{t})}{r_1(\boldsymbol{t})}}\right),
\end{equation}
with the respective intervals being $\theta \in [0, \pi]$ and $\varphi \in [0, 2\pi]$. In this mapping, the coordinates for the Pauli $Z$ eigenstates are $\ket{0}_L \mapsto (\theta, \varphi) = (0,0)$, and $\ket{1}_L \mapsto (\pi,0)$. Coordinates of stabilizer and magic states can be found in Appendix~\ref{appendix: bloch_vector_polar_coords}.

\subsection{Results}
The results can be found in~\cref{fig:histograms}. The states that produce high fidelity distillable magic states, such as $\ket{0}$ and $\ket{2}$, have projections close to $H$-type magic states, and as was mentioned in~\cite{Baragiola_2019, Calcluth_Sufficient_Condition_PRXQ_2024}, the square GKP code is biased towards the $H$-type magic states in the $X-Z$ plane for the vacuum state, unlike the hexagonal GKP code which is biased towards the $F$-type states. In fact, the histograms for most even-symmetric states with success probabilities greater than the vacuum bear a resemblance to~\cref{fig:histograms}(a) and~\cref{fig:histograms}(c).

What is interesting however, is the alternative interpretation suggested by the histograms. Namely, the states that perform \textit{worse} than the vacuum state, such as $\ket{1}, \ket{3}$ and $\ket{4}$, show high probabilities near the Pauli eigenstates. While $\ket{3}$ and $\ket{4}$ show high probabilities near the $H$-type magic states, there is still significant projection directly onto the stabilizer states. This suggests that the best states are those that actively avoid projection close to stabilizer states, regardless of their symmetries, rather than having good overlap with target magic states.

Unlike $\ket{0}$ and $\ket{2}$, the optimized Fock state $\ket{\Phi_{\mathrm{opt},5}}$ projects onto the $H$-type states in the $X-Y$ and $Y-Z$ plane, with relatively high probability. However there is almost zero probability of it being projected near the Pauli eigenstates. 

In the case of cat states, the odd $\ket{2\sqrt{\pi}, 0} - \ket{-2\sqrt{\pi}, 0}$ state situated on the GKP lattice is one of the few odd-symmetric states that has higher than vacuum success probabilities, and it is clear from its output state distribution that this state avoids the Pauli eigenstates. The distribution found in~\cref{fig:histograms}(g) forms a ring-like structure around the stabilizer states. 

Finally, the odd $\ket{\sqrt{\pi}, 0} - \ket{-\sqrt{\pi}, 0}$ state also placed on the GKP grid performs significantly worse than most other states examined in this paper, and its output state distribution sheds much light on this matter. This state projects onto all the Pauli eigenstates with pronounced probabilities, however it projects onto the $H$-type magic states with even higher probabilities. This supports the interpretation that states that perform better than the vacuum state are those that avoid projecting onto the Pauli eigenstates, as opposed to projecting onto the magic states themselves. 

We also note that the input states with even parity project onto the $H$-type magic states in the $X-Z$ plane whereas the states with odd-parity project onto the $H$-type states outside the $X-Z$ plane.

\section{Discussion} \label{sec: conc}

Here we analyzed the GKP error correction of simple non-Gaussian states in the context of the protocol introduced in Ref.~\cite{Baragiola_2019}. While originally applied only to thermal states, we have analytically and numerically shown that the extension of this protocol to Fock and cat states can offer significant improvements in the overall bosonic resources (i.e.,\ the total number of copies of the physical state) required to distill a desired logical magic state.  For example, we have found that the Fock state $\ket{2}$ consumes half as many resources as the vacuum, and that certain cat states can offer a three-fold reduction.  Such reductions offer an operational trade-off between bosonic non-Gaussianity and the ability to produce an encoded GKP magic state.  

That being said, this trade-off is subtle, with other non-Gaussian states offering only a marginal improvement or sometimes none at all.  In particular, we have found, perhaps surprisingly, that a bosonic state does not need to share any symmetry with the target magic state in order to yield an advantage.  For instance, we have found numerous cat states with their constituent coherent states not placed on either the logical or stabilizer lattice, that nonetheless perform better or have higher success probabilities than the truncated $\ket{+H}$ state or the vacuum.  As another example, the Fock states $\ket{1}$, $\ket{2}$, and $\ket{4}$ all have the same rotational symmetry as both the vacuum and the encoded state $\ket{+H}$, yet their performance for magic state distillation varies greatly.  In particular, $\ket{1}$ performs worse than the vacuum while $\ket{2}$ performs better, and, moreover, the Fock state $\ket{4}$ performs worse than both. This suggests that neither state symmetry nor photon parity acts as a guiding principle for optimizing this task; see also \cite{Grimsmo_Combes_Baragiola_PRX_2020}.  
 
Determining the precise relationship between bosonic non-Gaussianity, encoded magic, and distillation cost is a primary follow-up research direction.  Relatedly, throughout this work we used the 15-qubit Reed--Muller code as the concatenated distillation code, however, this was just one particular choice and not essential.  Other such codes will perform differently as an additional layer on top of the GKP code, and future work could be to determine the precise relationship between other distillation codes and bosonic non-Gaussianity. Moreover, achieving fault-tolerant universality through distillation presumes that logical Clifford operations can be implemented nearly perfectly, which is only justified when multimode GKP codes are considered, obtained for example through concatenation with qubit-level codes. The projection-based scheme used in~\cite{Baragiola_2019} and in the present work can be rephrased in this scenario but the resource analysis (in particular, the success probability) would need to be re-assessed. The same applies to the comparison between our proposed strategy and the direct preparation of magic states~\cite{Yamasaki_cost_reduced_2020} as well as the use of approximate logical stabilizer states~\cite{Hosseinynejad_Realistic_GKP_are_universal_PRL_2026}, which might scale poorly when applied to higher dimensional codes.

Finally, following \cite{Baragiola_2019}, we assumed throughout this work the ability to perform perfect GKP error-correction, which entails the creation of ideal GKP stabilizer states.  It will hence be of interest to further extend our analysis to the finite-energy regime. Indeed, while preparing this manuscript we became aware of the upcoming work by Bardales et al.~\cite{bardales_ferrini} in this direction.

\section*{Code availability}
The code and data used to produce all figures in this work are openly available at~\cite{gkpmsd_code}.

\begin{acknowledgments}
The authors acknowledge fruitful discussions with E.\ Bardales, C.\ Calcluth, T.\ Hillmann, L.\ García-Álvarez, A.\ Ferraro and G.\ Ferrini. S.D.\ is immensely grateful to T.\ Martinez, M.\ Uldemolins and P.\ Nair for guidance with the computational tools used in this work. The authors also acknowledge useful discussions with O.\ Hahn, M.\ Stafford and B.\ Baragiola. S.D.\ thanks V.\ Upreti for helpful comments on the manuscript. The authors acknowledge the use of Anthropic's Claude to refactor sections of the code, particularly those involving Riemann theta functions, from an earlier SageMath based implementation. S.D.\ acknowledges funding from the Hybrid Quantum Initiative (HQI) supported by France 2030 under the French National Research Agency (ANR) grant ANR-22-PNCQ0002.  J.D.\ and U.C.\ acknowledge funding from the European Union’s Horizon Europe Framework Programme (EIC Pathfinder Challenge project Veriqub) under Grant Agreement No.~101114899. F.A.\ acknowledges support from the French National Research Agency (ANR) through the Franco-German collaborative project BoLaCo (ANR-24-CE92-0076).
\end{acknowledgments}

\appendix
\onecolumngrid
\newpage
\section{Details of the numerical evaluations}

\subsection{Success probability \label{app:numerics_details_sprobs}}

The graphs seen in \cref{fig: cat_states}, \cref{fig:fock_states_sprobs}, \cref{fig:N_resources_cv_ratio} and \cref{fig:N_resources_fock_per_fid} were obtained by performing the double integral in~\cref{eqn:success_probability} at each minimum fidelity $F^*$. The code implementing everything described in this appendix, together with the data and the notebooks that generate each figure, is available at~\cite{gkpmsd_code}. A grid-based Riemann sum integration method was chosen because the domain of the integral is dependent on a threshold condition, and is therefore not smooth, which is a requirement for adaptive methods like SciPy's \textit{dblquad}. 100 minimum fidelity values were generated over the interval $[0,1]$, and this integration was carried out at each fidelity point. 

Since the integral required sweeping over the $[-\sqrt{\pi}, \sqrt{\pi})$ interval, this space was discretized into a uniform square grid, with the convergence of the integral depending on the number of divisions of this grid. This was implemented through a NumPy \textit{linspace} function with \textit{endpoint=False}: the syndrome domain is periodic with period $2\sqrt{\pi}$, so $-\sqrt{\pi}$ and $+\sqrt{\pi}$ label the same measurement outcome and only one of the two may be sampled. Then, the integral can be approximated as
\begin{equation}
    P_\mathrm{success} \approx \left(\sum_{t_q, t_p} \bar r_0(t_q, t_p)\times \chi(t_q, t_p) \right) \Delta t_q \Delta t_p, 
\end{equation}
given that the grid is uniform, $\Delta t_q = \Delta t_p = \Delta t$, which, for computational efficiency, becomes $(\Delta t)^2$. Here the grid spacing is
\begin{equation}
    \Delta t = \frac{2* \sqrt{\pi}}{N_\mathrm{divs}},
\end{equation}
where $N_\mathrm{divs}$ is the number of divisions per axis of the 2D grid. 

The Bloch vectors $(r_0(t_q, t_p),r_1(t_q, t_p),r_2(t_q, t_p),r_3(t_q, t_p))$, were computed using the relevant equations for cat or Fock states, at each point $(t_q, t_p)$, on this grid. For each Bloch vector, its \textit{best fidelity} was found by optimizing over the Clifford orbit using~\cref{eqn: best_fidelity_H}. This was done by calculating the dot product of the relevant state's Bloch vector, with the Bloch vectors of all 12 $H$-type magic states, with the largest dot product corresponding to the closest $H$-type magic state. 

Next, for each $F^*$, a Boolean mask was constructed by evaluating the pre-computed best fidelity array against the $F_{H,\mathrm{best}}(\boldsymbol{t}) \geq F^*$ condition. This mask was then applied to the probability of obtaining each $(t_q, t_p)$ outcome, $\bar r_0(t_q,t_p)$, and finally the entire probability array was summed over all the entries where the mask was \textit{True}, and multiplied by $(\Delta t)^2$, to obtain the final result. 

Two distinct notions of convergence are worth separating here. The normalization $\sum_{t_q, t_p} \bar r_0(\boldsymbol{t}) (\Delta t)^2$ carries no threshold mask, so its integrand is smooth and periodic over the syndrome domain; the corresponding Riemann sum is then exact to machine precision at any grid resolution, which we verified down to $N_\mathrm{divs} = 50$. The success probability curves at $F^* > 0$ converge more slowly, because the mask $\chi(\boldsymbol{t})$ restricts the domain of integration to a region whose boundary the grid resolves only to $O(1/N_\mathrm{divs})$. Measured against the $N_\mathrm{divs} = 400$ result, the curves agree to within $4 \times 10^{-2}$ at $N_\mathrm{divs} = 50$, $1.2 \times 10^{-2}$ at $N_\mathrm{divs} = 100$, and $3 \times 10^{-3}$ at $N_\mathrm{divs} = 200$. The curves in the aforementioned figures were computed with $N_\mathrm{divs} = 400$, i.e.\ a $400 \times 400$ grid.

To verify the numerical integrity, we performed the integral $\sum_{t_q, t_p} \bar r_0(\boldsymbol{t}) (\Delta t)^2$ for every state documented in this work, including all optimized cat states and Fock superpositions found by the procedures described in Appendix~\ref{app:numerics_details_optimization}. This normalization equals unity to within $2 \times 10^{-14}$ at $N_\mathrm{divs} = 400$, and to within $5 \times 10^{-15}$ already at $N_\mathrm{divs} = 50$. The discretization therefore conserves total probability exactly, and any deviation of the success probability curves from their converged values is attributable solely to the resolution of the threshold boundary. 

\subsection{Optimization over success probability \label{app:numerics_details_optimization}}

Optimization techniques enabled the discovery of several different superposition states, that were not only found to have higher success probabilities than the vacuum state but also lower distillation costs. Here we detail the specifics of how this optimization was executed. 

We maximized both cat and Fock state success probabilities with the SciPy global optimizer --- Differential Evolution. The differential evolution method~\cite{storn_differential_1997} is a derivative-free optimization algorithm that generates a population of candidate solutions and evolves them towards an optimal solution over multiple generations. For each candidate population, the cat or Fock state Bloch vector coefficients are found, the success probability is computed, and the optimization is carried out in one of two ways, either with the \textit{point} method, or with the \textit{area} method. 

In the case of the former, the success probability is computed at a single fidelity point, and this value passed to the maximizer. In the case of the area method, the success probability is computed for the entire interval of $F^*$ values from $[0,1]$, and the area under this curve passed to the maximizer. 

\subsubsection{Cat states}
We maximized the success probability of the cat state defined in~\cref{eqn:gen_cat_state} over both the amplitudes $\bar \gamma_c$ and the phase space coordinates, $q_0, p_0$. For the purpose of optimization, each complex amplitude was parametrized by its real and imaginary components, $a$ and $b$, then constructed into its complex form. Once a population was generated, the success probabilities were found with an interval grid of $400 \times 400$ for $\boldsymbol{t}$. In the case of the point method, the optimization was carried out for $\Lambda \in \{2, 3, 4, 5\}$ for $F^* \in \{ 0.94, 0.95, 0.96, 0.97, 0.98, 0.99\}$, these parameters were the inputs, and the outputs were the $q_0, p_0$ values of each of the coherent states, and their corresponding amplitudes, $\bar \gamma_c$. The optimizer ran for a maximum number of 50 iterations and a population size of 20 candidate solutions were generated at each iteration for both cases. In the case of the area method, the optimization was carried out for $\Lambda \in \{2, 3, 4, 5\}$.

\subsubsection{Fock states}
In the case of Fock states, we only needed to optimize over the complex amplitudes $\phi_i$ of the general finite superposition state defined in~\cref{eqn:fock_gen_superposition}. To significantly reduce the computational overhead we pre-computed the $f_0^{(n)}(t_q, t_p)$ and $f_1^{(n)}(t_q, t_p)$ values for Fock states $\ket{n} \in \{0,1,2,3,4\}$ over a grid of $400 \times 400$ discretized $\boldsymbol{t}$ values. Here the states were only optimized with the point method. The optimization was carried out for a superposition of 2, 3, 4, and 5 states, where a superposition of 2 states was $\Phi_\mathrm{opt,2} = \phi_0\ket{0} + \phi_1\ket{1}$, a superposition of 3 was $\Phi_\mathrm{opt,3} = \phi_0\ket{0} + \phi_1\ket{1} + \phi_2\ket{2}$, a superposition of 4 states was $\Phi_\mathrm{opt,3} = \phi_0\ket{0} + \phi_1\ket{1} + \phi_2\ket{2} + \phi_3\ket{3}$,and a superposition over all 5 states was $\Phi_\mathrm{opt,5} = \phi_0\ket{0} + \phi_1\ket{1} + \phi_2\ket{2} + \phi_3\ket{3} + \phi_4\ket{4}$.

\subsection{Optimization of the number of CV resources}
A similar optimization to the case of the success probabilities was done here, \cref{eqn:N_cv_resources} was minimized over to find superposition states which consumed the least amount of resources. This was only carried out on Fock states, therefore the pre-computed $f_0^{(n)}(t_q, t_p)$ and $f_1^{(n)}(t_q, t_p)$ values were utilized here as well, to obtain complex amplitudes $\phi_i$ for various superposition states.

\section{Truncated $\ket{+H_L}$ state calculations}

An alternate definition of the GKP code,~\cref{eqn: gkp_grimsmo_puri_alt}, allows for the codes states to be written as
\begin{equation}
    \ket{0_L} = \sum_{k,l} e^{-i \pi k l}\ket{2 k \alpha + l\beta} \quad \text{and} \quad \ket{1_L} = \sum_{k,l} e^{-i \pi (k l + l/2)}\ket{(2 k + 1) \alpha + l\beta},
\end{equation}
where $\alpha = \sqrt{\frac{\pi}{2}}$ and $\beta = i \sqrt{\frac{\pi}{2}}$~\cite{Grimsmo_2021}. Then, $\ket{+H_L}$ becomes
\begin{align}
    \ket{+H_L} & = \cos(\frac{\pi}{8})\ket{0_L} + \sin(\frac{\pi}{8})\ket{1_L}\\
    & = \cos(\frac{\pi}{8}) \sum_{k,l} e^{-i \pi k l}\ket{2 k \alpha + l\beta} + \sin (\frac{\pi}{8}) \sum_{k,l} e^{-i \pi (k l + \frac{l}{2})}\ket{(2 k + 1) \alpha + l\beta},
\end{align}
where $k,l$ can be substituted for with integer values to find different output states, the results of which are found in~\cref{tab:truncated_H_states}.
\begin{table}[h!]
    \centering
    \begin{tabular}{||c|c|c||}
        \hline
         $k$ & $l$ & State\\
         \hline
         0& 0 & $\cos(\frac{\pi}{8})\ket{0} + \sin(\frac{\pi}{8})\ket{\alpha}$\\
         0 & 1 & $\cos(\frac{\pi}{8})\ket{\beta} + i \sin(\frac{\pi}{8})\ket{\alpha + \beta}$\\
         0 & -1 & $\cos(\frac{\pi}{8})\ket{-\beta} - i \sin(\frac{\pi}{8})\ket{\alpha - \beta}$\\
         1 & 0 & $\cos(\frac{\pi}{8})\ket{2\alpha} + \sin(\frac{\pi}{8})\ket{2\beta}$\\
         1 & 1 & $-\cos(\frac{\pi}{8})\ket{2\alpha + \beta} + i \sin(\frac{\pi}{8})\ket{3\alpha + \beta}$\\
         1 & -1 & $-\cos(\frac{\pi}{8})\ket{2\alpha - \beta} + i \sin(\frac{\pi}{8})\ket{3\alpha - \beta}$\\
         \hline
         \hline
    \end{tabular}
    \caption{$\ket{+H}_L$ states for various combinations of $k$ and $l$.}
    \label{tab:truncated_H_states}
\end{table}
We set $k = 0$ and $l = \{0,1\}$ to find a four-legged cat state
\begin{equation}
    \ket{+H_L} = \cos(\frac{\pi}{8})\ket{0} + \sin(\frac{\pi}{8})\ket{\alpha} + \cos(\frac{\pi}{8})\ket{\beta} + i \sin(\frac{\pi}{8})\ket{\alpha + \beta},
\end{equation}
using the notation established in the main paper, we re-write the above in terms of the positions in the two quadratures, so $q = \sqrt{2}\alpha$ and $p = \sqrt{2}\Im{\beta}$, then the four-legged truncated $H$ state seen in~\cref{fig:cat_trunc_h_state} is
\begin{equation}
    \ket{\Gamma_4} = \cos(\frac{\pi}{8})\ket{0,0} + \sin(\frac{\pi}{8})\ket{\sqrt{\pi}} + \cos(\frac{\pi}{8})\ket{0,\sqrt{\pi}} - i\sin(\frac{\pi}{8})\ket{\sqrt{\pi}, \sqrt{\pi}}.
\end{equation}
To obtain the six-legged truncated $\ket{+H_L}$ state seen in~\cref{fig:cat_trunc_h_state}, we set $k = 0$, and $l = \{-1,0,1\}$ to find
\begin{equation}
    \ket{\Gamma_6} = \cos(\frac{\pi}{8})\left (\ket{0,\sqrt{\pi}} + \ket{0,0} + \ket{0, -\sqrt{\pi}}\right) + \sin(\frac{\pi}{8})(-i \ket{\sqrt{\pi}, \sqrt{\pi}} + \ket{\sqrt{\pi}} + i\ket{\sqrt{\pi}, -\sqrt{\pi}}).
\end{equation}
Other truncated $H$ states were found for different combinations of $k,l$, however they did not result in better success probabilities.

\section{Optimized states\label{appendix: opt_state_descriptions}}
Here we note the descriptions of the states shown in Figures~\ref{fig:cat_trunc_h_state}--\ref{fig:cat_opt_positions}. We start with $\ket{\Gamma_\mathrm{opt,4'}}$ which was optimized over $F^* = 0.95$,
 \begin{equation}
 \begin{split}
    \ket{\Gamma_\mathrm{opt,4'}} & = (-0.497+0.347i)\ket{0.3\sqrt{\pi}, 0.168\sqrt{\pi}} + (0.536-0.058i)\ket{1.229\sqrt{\pi}, -1.184\sqrt{\pi}} \\
    &+ (0.627+0.412i)\ket{1.885\sqrt{\pi}, -1.288\sqrt{\pi}} + (0.435-0.138i)\ket{1.19\sqrt{\pi}, 0.348\sqrt{\pi}}.\\
\end{split}
\end{equation}

Then we list the states from~\cref{fig:cat_states_opt_sprobs}, optimized over $F^* = 0.99$,
\begin{equation}
    \ket{\Gamma_\mathrm{opt,2}} = (0.694 + 0.896i)\ket{-1.057\sqrt{\pi}, -1.865\sqrt{\pi}} + (-0.872 - 0.855i)\ket{-1.075 \sqrt{\pi}, -0.051\sqrt{\pi}},
\end{equation}
\begin{equation}
\begin{split}
    \ket{\Gamma_\mathrm{opt,3}} & = (-0.624-0.994i)\ket{0.124\sqrt{\pi}, -2.816\sqrt{\pi}} + (-0.494-0.208i)\ket{ -1.643\sqrt{\pi}, 0.088\sqrt{\pi}}\\
    & + (0.325-0.779i)\ket{-0.555\sqrt{\pi}, -0.861\sqrt{\pi}},\\
\end{split}
\end{equation}
\begin{equation}
\begin{split}
    \ket{\Gamma_\mathrm{opt,4}} & = (0.99-0.224i)\ket{1.059\sqrt{\pi}, 0.915\sqrt{\pi}} + (-0.05-0.951i)\ket{-0.522\sqrt{\pi}, 0.854\sqrt{\pi}}\\
    & + (0.179+0.916i)\ket{1.004\sqrt{\pi}, 0.659\sqrt{\pi}} + (-0.776+0.619i)\ket{ -0.851\sqrt{\pi}, -0.947\sqrt{\pi}},
\end{split}
\end{equation}
\begin{equation}
\begin{split}
    \ket{\Gamma_\mathrm{opt,5}} & = (-0.636-0.095i)\ket{-0.152\sqrt{\pi}, 1.474\sqrt{\pi}} + (-0.388-0.272i)\ket{0.984\sqrt{\pi}, 0.96\sqrt{\pi}}\\
    & + (-0.081-0.922i)\ket{ 1.364\sqrt{\pi}, 2.474\sqrt{\pi}} + (-0.868-0.412i)\ket{ -0.428\sqrt{\pi}, 0.312\sqrt{\pi}}\\
    & + (-0.912+0.151i)\ket{ -0.582\sqrt{\pi}, -1.338\sqrt{\pi}}.\\
\end{split}
\end{equation}

\section{Fock state overlaps with the GKP logical operators}
Here we show the equations for the overlaps of Fock states with Pauli $I$, $Y$ and $Z$:
\begin{equation}
\begin{split}
    \bar r_0^{(n)} (\boldsymbol{t}) = \underbrace{\bra{n}V^\dagger(-\boldsymbol{t})\ket{0_L}}_{{f_0^{(n)}(\boldsymbol{t})}^*}\cdot \underbrace{\bra{0_L}V(-\boldsymbol{t})\ket{n}}_{f_0^{(n)}(\boldsymbol{t})} + \underbrace{\bra{n}V^\dagger(-\boldsymbol{t})\ket{1_L}}_{{f_1^{(n)}(\boldsymbol{t})}^*} \cdot \underbrace{\bra{1_L}V(-\boldsymbol{t})\ket{n}}_{f_1^{(n)}(\boldsymbol{t})},\\
\end{split}
\end{equation}
\begin{equation}
\begin{split}
    \bar r_2^{(n)} (\boldsymbol{t}) = -i\left(\underbrace{\bra{n}V^\dagger(-\boldsymbol{t})\ket{0_L}}_{{f_0^{(n)}(\boldsymbol{t})}^*}\cdot \underbrace{\bra{1_L}V(-\boldsymbol{t})\ket{n}}_{f_1^{(n)}(\boldsymbol{t})}\right) + i\left( \underbrace{\bra{n}V^\dagger(-\boldsymbol{t})\ket{1_L}}_{{f_1^{(n)}(\boldsymbol{t})}^*} \cdot \underbrace{\bra{0_L}V(-\boldsymbol{t})\ket{n}}_{f_0^{(n)}(\boldsymbol{t})}\right),\\
\end{split}
\end{equation}
\begin{equation}
\begin{split}
    \bar r_3^{(n)} (\boldsymbol{t}) = \underbrace{\bra{n}V^\dagger(-\boldsymbol{t})\ket{0_L}}_{{f_0^{(n)}(\boldsymbol{t})}^*}\cdot \underbrace{\bra{0_L}V(-\boldsymbol{t})\ket{n}}_{f_0^{(n)}(\boldsymbol{t})} - \underbrace{\bra{n}V^\dagger(-\boldsymbol{t})\ket{1_L}}_{{f_1^{(n)}(\boldsymbol{t})}^*} \cdot \underbrace{\bra{1_L}V(-\boldsymbol{t})\ket{n}}_{f_1^{(n)}(\boldsymbol{t})}.\\
\end{split}
\end{equation}

\section{The overlap equations for Fock states \label{app:fock_states_rmus}}
The overlap, $\fockgk$, for states $\ket{0}, \ket{2}, \ket{3}$ and $\ket{4}$. These equations are derived using the method detailed in \cref{sec:bloch_vec_calcs_fock}.
\subsection{Fock state 0 coefficients}
\begin{equation}
    \fockgk[0] = \frac{1}{2 \sqrt[4]{\pi}} G_{\bar 0}(\boldsymbol{t}) \Theta(z_{\bar n}(\boldsymbol{t}), \Omega_{\bar n}).
\end{equation}
Note that this overlap can also be obtained by substituting with $\alpha_c = 0$ in ~\cref{eqn:cat_gk_expanded}.
\subsection{Fock state 2 coefficients}
To find the overlaps for $\ket{2}$, we continue from $ \fockgk$
\begin{equation}
    \fockgk[2]= \frac{1}{2 \sqrt[4]{\pi}} \sum_l e^{-i t_p (2l + k)\sqrt{\pi}}\underbrace{{}_q\bra{(2l + k)\sqrt{\pi} +t_q} \ket{2}}_{\psi_2(q = (2l + k)\sqrt{\pi} +t_q)},
\end{equation}
where now the wavefunction is $\psi_2(q) = \frac{1}{\sqrt{8}} \frac{1}{\pi^{1/4}} e^{-\frac{q^2}{2}} (4q^2 - 2)$. Then we obtain
\begin{equation}
\begin{split}
    \fockgk[2] & = \frac{1}{2 \sqrt[4]{\pi}} \sum_l e^{i (-t_p) (2l + k)\sqrt{\pi}} \left [ \frac{1}{2\sqrt{2}} \frac{1}{\pi^{1/4}} \exp\left[-2l^2 \pi - 2lk\pi - \frac{k^2 \pi}{2} - 2l\sqrt{\pi}t_q - k\sqrt{\pi}t_q - \frac{t_q^2}{2}\right] (4((2l + k)\sqrt{\pi} + t_q)^2 - 2) \right]\\
    & =  \frac{1}{2 \sqrt[4]{\pi}} \underbrace{\frac{\exp\left[ -i t_p k\sqrt{\pi} - \frac{k^2 \pi}{2} - k\sqrt{\pi}t_q - \frac{t_q^2}{2} \right]}{2\sqrt{2}\pi^{1/4}}}_{\text{Gaussian term } G_{\bar 2}(\boldsymbol{t})}\underbrace{\underbrace{\sum_l\exp\left [-2l^2 \pi - 2lk\pi - 2l\sqrt{\pi} t_q - i t_p 2l \sqrt{\pi} \right]}_{\text{Riemann theta }\Theta( z_{\bar n}(\boldsymbol{t}), \Omega_{\bar n})} 4((2l + k)\sqrt{\pi} + t_q)^2 - 2)}_{\text{derivatives of the Riemann theta}}\\
    & = \frac{1}{2 \sqrt[4]{\pi}} G_{\bar 2}(\boldsymbol{t}) \sum_l \exp[2\pi i \left( \frac{1}{2}(2 i )l^2 + l \left( ik - \frac{t_p}{\sqrt{\pi}} + \frac{i t_q}{\sqrt{\pi}}\right) \right)] (16\pi l^2 + 16l k \pi + 16 l \sqrt{\pi} t_q + 4k^2 \pi + 8 k \sqrt{\pi} t_q + 4 t_q^2 - 4).
\end{split}
\end{equation}
Finally, we can write the above as
\begin{equation}
    \fockgk[2] = \frac{1}{2 \sqrt[4]{\pi}} G_{\bar 2}(\boldsymbol{t}) \left(A_2 \frac{d^2\Theta(z_{\bar n}(\boldsymbol{t}), \Omega_{\bar{n}})}{dz^2} + A_1 \frac{d\Theta(z_{\bar n}(\boldsymbol{t}), \Omega_{\bar{n}})}{dz} + A_0 \Theta(z_{\bar n}(\boldsymbol{t}), \Omega_{\bar{n}})\right)
\end{equation}
where $A_2 = \frac{-4}{\pi}$, $A_1 = -8 i \left(k + \frac{t_q}{\sqrt{\pi}}\right)$ and $A_0 = (4k^2 \pi + 8k\sqrt{\pi} t_q + 4t_q^2 - 2)$.

\subsection{Fock state 3 coefficients}
\begin{equation}
    \fockgk[3] =  \frac{1}{2 \sqrt[4]{\pi}} G_{\bar 3}(\boldsymbol{t}) \left (A_3\frac{d^3\Theta(z_{\bar n}(\boldsymbol{t}), \Omega_{\bar n})}{dz^3} + A_2 \frac{d^2\Theta(z_{\bar n}(\boldsymbol{t}), \Omega_{\bar n})}{dz^2} + A_1 \frac{d\Theta(z_{\bar n}(\boldsymbol{t}), \Omega_{\bar n})}{dz} + A_0 \Theta(z_{\bar n}(\boldsymbol{t}), \Omega_{\bar n}) \right )
\end{equation}
where $A_3 = \frac{8 i}{\pi^{3/2}}$, $A_2 = \frac{-24 k}{\sqrt{\pi}} - \frac{24 t_q}{\pi}$, $A_1 = \frac{12 i}{\sqrt{\pi}} - 24 i k^2 \sqrt{\pi} - 48 i k t_q - \frac{24 i t_q^2}{\sqrt{\pi}}$, $A_0 = -12 k\sqrt{\pi} + 8 k^3 \pi^{3/2} - 12 t_q + 24 k^2 \pi t_q + 24 k \sqrt{\pi} t_q^2 + 8 t_q^3$.
\subsection{Fock state 4 coefficients}
\begin{equation}
\begin{split}
    \fockgk[4] = \frac{1}{2 \sqrt[4]{\pi}} G_{\bar 4}(\boldsymbol{t}) & \left (A_4\frac{d^4\Theta(z_{\bar n}(\boldsymbol{t}), \Omega_{\bar n})}{dz^4} +  A_3 \frac{d^3\Theta(z_{\bar n}(\boldsymbol{t}), \Omega_{\bar n})}{dz^3} + A_2 \frac{d^2\Theta(z_{\bar n}(\boldsymbol{t}), \Omega_{\bar n})}{dz^2} \right . \\
    & \left . + A_1 \frac{d\Theta(z_{\bar n}(\boldsymbol{t}), \Omega_{\bar n})}{dz} + A_0\Theta(z_{\bar n}(\boldsymbol{t}), \Omega_{\bar n})\right )\\
\end{split}
\end{equation}
where $A_4 = \frac{16}{\pi^2}$, $A_3 = \frac{64 i k}{\pi} + \frac{64 i t_q}{\pi^{3/2}}$, $A_2 = -96k^2 + \frac{48}{\pi} - \frac{192 k t_q}{\sqrt{\pi}} - \frac{96 t_q^2}{\pi}$, $A_1 = 96 i k - 64 i k^3 \pi + \frac{96 i t_q}{\sqrt{\pi}} - 192 i k^2 \sqrt{\pi} t_q - 192 i k t_q^2 - \frac{64 i t_q^3}{\sqrt{\pi}}$ and $A_0 = 12 - 48 k^2 \pi + 16k^4 \pi^2 - 96 k\sqrt{\pi} t_q + 64 k^3 \pi^{3/2} t_q - 48 t_q^2 + 96 k^2 \pi t_q^2 + 64 k \sqrt{\pi} t_q^3 + 16 t_q^4$.

\section{Bloch vectors of magic states and stabilizer states and their polar coordinates. \label{appendix: bloch_vector_polar_coords}}
\begin{table}[!h]
    \centering
\begin{tabular}{|||c|c|c||c|c|c|||}
    \hline
     State & Bloch vector & Polar co-ordinate $(\theta, \varphi)$ & State & Bloch vector & Polar co-ordinate $(\theta, \varphi)$\\
     \hline
     \hline
     $\ket{+}$ & $(1,0,0)$ & ($\frac{\pi}{2}, 0$) & $\ket{-}$ & $(-1,0,0)$ & ($\frac{\pi}{2}, \pi$)\\
     $\ket{+_i}$ & $(0,1,0)$ & ($\frac{\pi}{2}, \frac{\pi}{2}$) & $\ket{-_i}$ & $(0,-1,0)$ & ($\frac{\pi}{2}, \frac{3\pi}{2}$)\\
     $\ket{0}$ & $(0,0,1)$ & ($0,0$) & $\ket{1}$ & $(0,0,-1)$ & ($\pi,0$)\\
     \hline
     $\ket{+H}$ & ($\frac{1}{\sqrt{2}}, 0, \frac{1}{\sqrt{2}}$) & ($\frac{\pi}{4}, 0$) & $\ket{-H}$ & ($\frac{-1}{\sqrt{2}}, 0, \frac{-1}{\sqrt{2}}$) & ($\frac{3\pi}{4}, \pi$)\\
     $Z\ket{+H}$ & ($\frac{-1}{\sqrt{2}}, 0, \frac{1}{\sqrt{2}}$) & ($\frac{\pi}{4}, \pi$) & $X\ket{+H}$ & ($\frac{1}{\sqrt{2}}, 0, \frac{-1}{\sqrt{2}}$) & ($\frac{3\pi}{4}, 0$)\\
     \hline
     $\ket{+H_y}$ & ($0, \frac{1}{\sqrt{2}}, \frac{1}{\sqrt{2}}$) & ($\frac{\pi}{4}, \frac{\pi}{2}$) & $\ket{-H_y}$ & ($0, \frac{-1}{\sqrt{2}}, \frac{-1}{\sqrt{2}}$) & ($\frac{3\pi}{4}, \frac{3\pi}{2}$)\\
     $Z\ket{+H_y}$ & ($0, \frac{-1}{\sqrt{2}}, \frac{1}{\sqrt{2}}$) & ($\frac{\pi}{4}, \frac{3\pi}{2}$) & $X\ket{+H_y}$ & ($0, \frac{1}{\sqrt{2}}, \frac{-1}{\sqrt{2}}$) & ($\frac{3\pi}{4}, \frac{\pi}{2}$)\\
     \hline
     $\ket{+H_{xy}}$ & ($\frac{1}{\sqrt{2}}, \frac{1}{\sqrt{2}}, 0$) & ($\frac{\pi}{2}, \frac{\pi}{4}$) & $\ket{-H_{xy}}$ & ($\frac{-1}{\sqrt{2}}, \frac{-1}{\sqrt{2}}, 0$) & ($\frac{\pi}{2}, \frac{5\pi}{4}$)\\
     $Z\ket{+H_{xy}}$ & ($\frac{-1}{\sqrt{2}}, \frac{1}{\sqrt{2}}, 0$) & ($\frac{\pi}{2}, \frac{3\pi}{4}$) & $Y\ket{+H_{xy}}$ & ($\frac{-1}{\sqrt{2}}, \frac{1}{\sqrt{2}}, 0$) & ($\frac{\pi}{2}, \frac{7\pi}{4}$)\\
     \hline
     $\ket{+F}$ & ($\frac{1}{\sqrt{3}}, \frac{1}{\sqrt{3}}, \frac{1}{\sqrt{3}}$) & ($\arccos(\frac{1}{\sqrt{3}}), \frac{\pi}{4}$) & $\ket{-F}$ & ($\frac{-1}{\sqrt{3}}, \frac{-1}{\sqrt{3}}, \frac{-1}{\sqrt{3}}$) & ($\arccos(\frac{-1}{\sqrt{3}}), \frac{5\pi}{4}$)\\
     $Z\ket{+F}$ & ($\frac{-1}{\sqrt{3}}, \frac{-1}{\sqrt{3}}, \frac{1}{\sqrt{3}}$) & ($\arccos(\frac{1}{\sqrt{3}}), \frac{5\pi}{4}$) & $Y\ket{+F}$ & ($\frac{-1}{\sqrt{3}}, \frac{1}{\sqrt{3}}, \frac{-1}{\sqrt{3}}$) & ($\arccos(\frac{-1}{\sqrt{3}}), \frac{3\pi}{4}$)\\
     $X\ket{+F}$ & ($\frac{1}{\sqrt{3}}, \frac{-1}{\sqrt{3}}, \frac{-1}{\sqrt{3}}$) & ($\arccos(\frac{-1}{\sqrt{3}}), \frac{7\pi}{4}$) & $X\ket{-F}$ & ($\frac{-1}{\sqrt{3}}, \frac{1}{\sqrt{3}}, \frac{1}{\sqrt{3}}$) & ($\arccos(\frac{1}{\sqrt{3}}), \frac{3\pi}{4}$)\\
     $Y\ket{-F}$ & ($\frac{1}{\sqrt{3}}, \frac{-1}{\sqrt{3}}, \frac{1}{\sqrt{3}}$) & ($\arccos(\frac{1}{\sqrt{3}}), \frac{7\pi}{4}$) & $Z\ket{-F}$ & ($\frac{1}{\sqrt{3}}, \frac{1}{\sqrt{3}}, \frac{-1}{\sqrt{3}}$) & ($\arccos(\frac{-1}{\sqrt{3}}), \frac{\pi}{4}$)\\
     \hline
\end{tabular}
    \caption{The Bloch vectors and polar coordinates for the $H$-type and $F$-type magic states and stabilizer states. The states $\ket{+H_y}$ and $\ket{+H_{xy}}$ are generated from $\ket{+H}$ as follows:
$ \ket{+H_y} = S \ket{+H} $
 $\ket{+H_{xy}} = S H S \ket{+H}$, where the phase gate is defined as $S = \begin{pmatrix} 1 & 0 \\ 0 & i \end{pmatrix}$. }
    \label{tab: bloch_vec_polar_coords_table}
\end{table}

\twocolumngrid
\bibliography{references}

\end{document}